\documentclass[final,3p,times]{elsarticle}

   \newcommand{\exclude}[1]{}
\newcommand{\be}{\begin{eqnarray}}
\newcommand{\ee}{\end{eqnarray}}

\usepackage{graphicx,amsmath,amssymb,bm}

\usepackage{psfrag}
 \usepackage{feynmp}
 \usepackage{hyperref}

\journal{Physics of the Dark Universe}

\newcommand{\beq}{\begin{equation}}
\newcommand{\eeq}{\end{equation}}

\def\ra{\rangle}
\def\la{\langle}

\begin{document}

\begin{frontmatter}

\title{ Solar Flares  and the Axion Quark Nugget Dark Matter Model} 
\author{Ariel  Zhitnitsky}
\address{ Department of Physics \& Astronomy, University of British Columbia, 
Vancouver, B.C. V6T 1Z1, Canada} 
 \begin{abstract}
We advocate the idea that the  nanoflares conjectured by Parker long ago to resolve  the corona heating problem, 
may also trigger the  larger solar flares. The arguments are based on the model where 
emission of extreme ultra violet (EUV) radiation and  soft x-rays from the Sun are powered  externally by incident dark matter  
particles  within the Axion Quark Nugget (AQN) 
Dark Matter Model.   The corresponding annihilation events of the AQNs with the solar material are identified with nanoflares.  This    model was originally invented as a natural explanation of  the 
observed ratio $\Omega_{\rm dark} \sim   \Omega_{\rm visible}$ when the DM and visible matter densities assume the same order of magnitude  values. This model gives a very reasonable intensity of EUV radiation without  adjustments of any parameters 
in the model. When the same nuggets enter the regions with high magnetic field they  serve as the triggers igniting  the magnetic reconnections which eventually may lead to much larger   flares. 

 Technically, the magnetic reconnection is ignited due to the shock wave which inevitably    develops  when the dark matter nuggets enter the solar atmosphere with velocity $v_{\rm AQN}\sim ~600~ {\rm km/s}$ which is much higher than the speed of sound $c_s$, such that the Mach number $M=v_{\rm AQN}/c_s\gg 1$. These shock waves generate very strong  and very short impulses expressed in terms of  pressure $\Delta p/p\sim M^2$  and temperature $\Delta T/T\sim M^2$   in vicinity of   (would be) magnetic reconnection area. 
 We find that this mechanism is consistent with   x -ray observations as well as  with observed jet like morphology of the initial stage of the flares. The mechanism is also consistent with the observed scaling of the flare  distribution   $dN\sim W^{-\alpha}dW$ as a function of the flare's energy $W$.  We also speculate that the same nuggets  may trigger the sunquakes which are known to be correlated with large flares.  
 
 \end{abstract}

\end{frontmatter}
 
\section{Introduction}

A variety of anomalous solar phenomena still defy conventional theoretical understanding. For example, the detailed physical processes that heat the outer atmosphere of the Sun to $10^6$K remain  a major open issue in astrophysics, see e.g. \cite{Klimchuk:2005nx} for review and references on the original results.    This persisting puzzle is characterized by the following observed anomalous behaviour of the sun:
the quiet Sun emits an extreme ultra violet (EUV) radiation  with a photon energy of order of hundreds of {\rm eV  } which cannot be explained in terms of  any conventional astrophysical phenomena;  the total energy output of the corona is quite small, see (\ref{estimate}) below. However, it    never drops to zero as time evolves. 
To be more specific,  the  total intensity of the  observed EUV and soft x-ray radiation (averaged over time)   can be estimated as follows,
\be
\label{estimate}
   L_{\odot ~  (\rm from ~Corona)}  \sim 10^{30}\cdot \frac{\rm GeV}{\rm  s} \sim 10^{27}  \cdot  \frac{\rm erg}{\rm  s}.
 \ee
  which represents about $(10^{-7}-10^{-6})$ fraction of the  solar luminosity. 

At the transition region, the (quiet Sun) temperature continues to rise very steeply  until it reaches  a few $10^{6}$~K, i.e., being a few 100 times hotter above the underlying  photosphere, and this within an atmospheric layer thickness of only 100 km or even much less. 
Therefore, after several decades of research, it may be that the answer on these (and many others related)   questions  lies in a new physics.  

\begin{figure}
	\centering
	 	 		\includegraphics[width=0.7\textwidth]{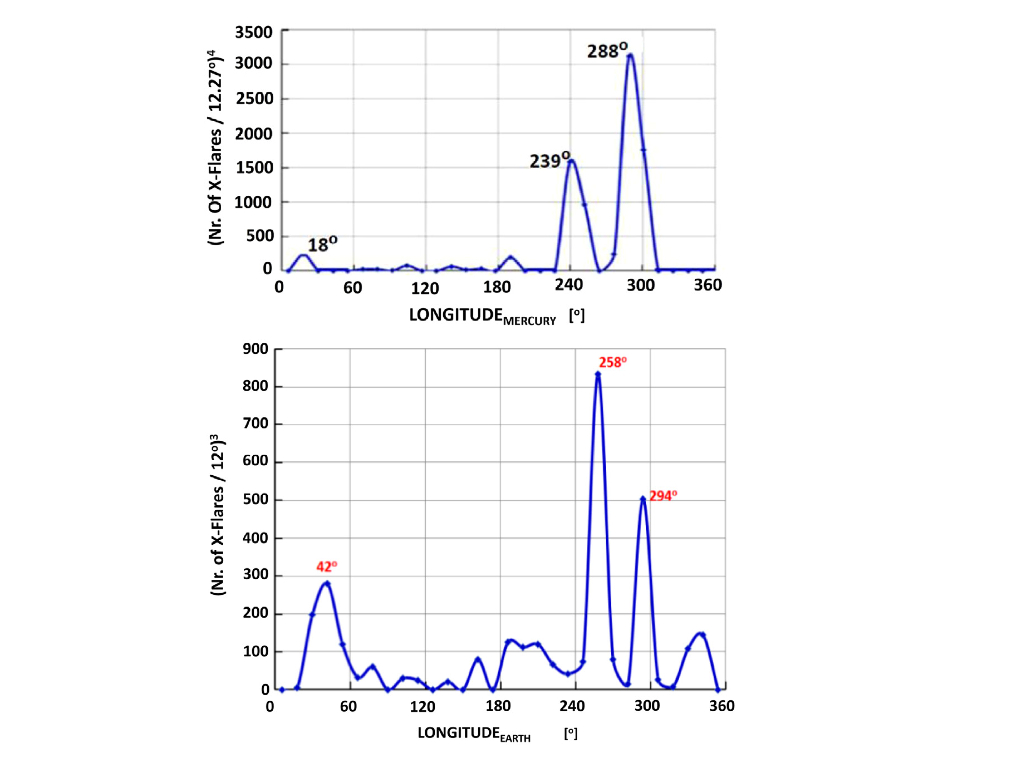}
	 	\caption{ \label{Earth-Mercury}The ``multiplication spectrum" is defined as $Y(J)\equiv \Theta_1(J)\cdot \Theta_2(J)\cdot \Theta_3(J)\cdot \Theta_4(J)$,
		where $\Theta$ is the deflection angle related to the gravitational lensing analysis.  
		The subscript (1-4) denotes the four solar cycles (1975-1986, 1986-1997, 1997-2009, 2009-2014) and $J$ denotes the bin number with widths ($6^{\circ}, 12^{\circ}, 16^{\circ}$).
		 The plots are taken from \cite{Zioutas}.}
		\end{figure}

\begin{figure}
	\centering
	 		\includegraphics[width=0.8\textwidth]{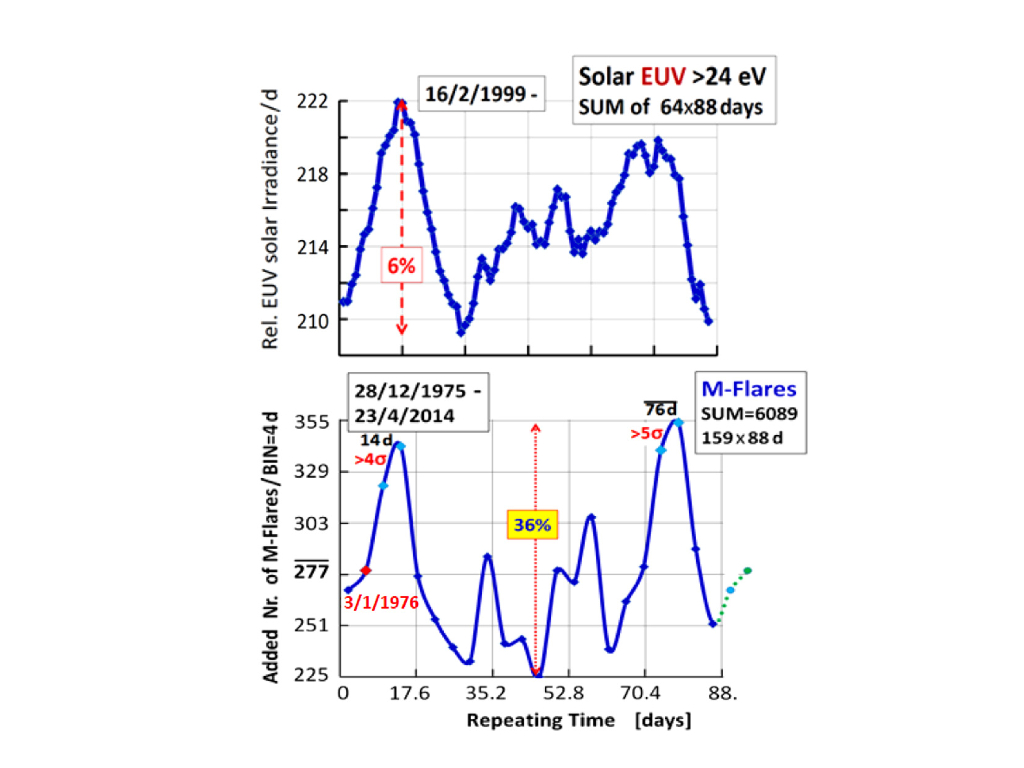}
			 	\caption{\label{EUV}This plot demonstrates  the connection  between  the relative EUV radiation (top) and the number of M-flares during the same period of time (bottom).  The BIN size is 1 day for the EUV data and 4 days for M-flares analysis. The plots are taken from \cite{Zioutas}.}
\end{figure}

It was precisely the main subject  of ref.\cite{Zioutas} where it was advocated  that a number of highly unusual phenomena  (including, but not limited to  the EUV radiation) observed in solar atmosphere might be related  to the gravitational lensing of ``invisible" streaming matter towards the Sun. 
The main argument of ref.\cite{Zioutas} is based on analysis of a number of different   correlations between 
 the relative orientations of the  Sun and its planets on the   one hand   and  the frequency of the observed flares on  the other hand  of the analysis.
 As an example of the observed correlations we reproduce on Fig.\ref{Earth-Mercury} some sample  plots 
 for the so-called ``multiplication spectrum" from ref. \cite{Zioutas} where 
 the frequency of occurring of the X-flares  is shown as a function of the Mercury (on the top) and the Earth (on the bottom) heliocentric longitude.  
 We refer to the original paper \cite{Zioutas} for the  specific discussions, definitions  and the details  on the data analysis. In this Introduction the only comment we would like to make is that one should not  expect    any correlations between the X flare occurrences   and the position of the planets. Nevertheless,    Fig.\ref{Earth-Mercury} obviously demonstrates that this naive expectation is not quite correct, and that the enhancement is happening around the same heliocentric longitude for Mercury and Earth in spite of the fact that periods of the Earth and Mercury are very different ($T_{\rm Mercury}= 87.969$ days) and they appear at this specific heliocentric longitude when the enhancement occurs, in general,  at different moments.

One should comment here that ``the solar corona heating problem" includes a number of elements which are hard to explain using a  conventional framework. In particular, the hot corona cannot be in equilibrium with the $\sim 300$ times cooler solar surface (violating thus the second law of thermodynamics  \cite{Wolfson}).  In order to maintain the quiet Sun high temperature corona, some non-thermally supplied energy must be dissipated in the upper atmosphere \cite{Alexander}. There are many other problems which are nicely stated  in review article \cite{Aschwanden} as follows
 ``everything above the photosphere...would not be there at all". In particular, the observed x-rays in 1-10 keV energy range in the non-flaring Sun are hard to  explain using traditional solar physics \cite{LZ:2003}. 
 
 It has been recently argued in \cite{Zhitnitsky:2017rop}  that the dark matter AQNs might be   the source of the heating of outer atmosphere of the Sun and, therefore,  might be directly responsible for  the observed EUV radiation and heating the corona. 
 Recently, this proposal has received a strong numerical support \cite{Raza:2018gpb}. 
 We review the basic arguments supporting this proposal in Section  \ref{AQN-nanoflares}. The only comment we would like to make here is that this proposal simultaneously 
   resolves the  two problems mentioned above: ``solar corona heating" problem and  mysterious and unexpected     correlation  of the solar activity with position of its planets. 
   \exclude{In other words, the dark matter AQNs might be   the source of the heating of outer atmosphere of the Sun and, therefore,  might be directly responsible for  the observed EUV radiation. In addition, these AQNs  represent  the ``invisible matter"  conjectured in \cite{Zioutas}.
    }
    
     To recapitulate:   this proposal links the EUV radiation and 
 occurrences of the  X, M flares  as these two apparently distinct phenomena in fact intimately  related   as they  obviously accompany each other according to   Fig.\ref{EUV}, and they {\it both} are correlated with positions of the planets\footnote{We should comment here that  a deep  relation between these two distinct phenomena should not be confused with equal-time correlation between the two. The connection which is shown on Fig.\ref{EUV} has pure statistical interpretation. Essentially it  demonstrates  that  the higher  intensity of the ``invisible" streaming matter toward the Sun (averaged over many solar cycles) generates a higher intensity for the EUV radiation. The same 
 increase of the the ``invisible" matter flux also leads  (again, averaged over many solar cycles)  to a   higher frequency of the  flare occurrences. }  as argued in  \cite{Zioutas}.
   
     It turns out that if one estimates    the extra energy being produced within the AQN dark matter scenario   one obtains  the total extra energy $\sim 10^{27}{\rm erg}/{\rm  s}$    which 
precisely reproduces  (\ref{estimate})   for  the   observed EUV and soft x-ray intensities  \cite{Zhitnitsky:2017rop}. 
The numerical Monte Carlo simulations \cite{Raza:2018gpb} fully support  the estimate  (\ref{estimate}). Furthermore,
the numerical analysis   \cite{Raza:2018gpb} also  shows that the energy injection occurs precisely in the vicinity of the transition region at the altitude $\sim 2000$ km. This confirms the order of magnitude estimate  \cite{Zhitnitsky:2017rop} that the emission will be mostly in form of the EUV and soft x-rays.

One should emphasize  that the production  of  extra energy  is   expressed exclusively  in terms of known  dark matter density $\rho_{\rm DM} \sim 0.3~ {\rm GeV cm^{-3}}$ and dark matter  velocity $v_{\rm DM}\sim 10^{-3}c $ surrounding the Sun  
 without adjusting any  parameters of the AQN model, see  section \ref{AQN-nanoflares} below with relevant comments.  We  interpreted   this ``numerical coincidence"    as an additional argument  supporting  our  proposal  that    the heating of the corona and the chromosphere      is originated form the AQNs entering the solar atmosphere from outer space.

 The main purpose of the present work is to present a very specific mechanism which may provide a hint on  how these two naively  distinct phenomena nevertheless might be  closely related to each other.
 This deep relation between these two distinct effects   may shed some light on   the   nature   of    the dark matter which (within the AQN paradigm) is the source for both  phenomena, the EUV  radiation and the flare's activity.  Therefore, increase of the DM flux  should lead to  
 some increase of  the EUV    radiation along with the higher frequency of  the flare's occurrences when averaged over long period of time.     
  
    One should emphasize that the AQN dark matter model 
  was  originally proposed  long ago  \cite{Zhitnitsky:2002qa} 
   without any connection to the solar system and the EUV. Rather,     the main motivation to develop  the AQN dark matter model   was  to   explain in a very natural way the observed   similarity between the visible and dark matter densities 
   in the present Universe, i.e. $\Omega_{\rm dark}\sim \Omega_{\rm visible}$.
   \exclude{, see short overview of the AQN model in the next section \ref{sec:QNDM}.}

 The paper is organized as follows. In next section \ref{sec:QNDM} we overview the AQN model by paying special attention to the astrophysical and cosmological consequences of this specific dark matter  model. In section \ref{AQN-nanoflares} we highlight  the basic arguments of refs.  \cite{Zhitnitsky:2017rop,Raza:2018gpb}
 advocating the idea that the annihilation events of the antinuggets with the solar material can be interpreted as the {\it nanoflares} conjectured  by Parker      long ago. In the   section \ref{flares} we argue that the same AQNs entering the region with the large magnetic field may spark the magnetic reconnection leading to very large {\it flares}. In other words, we want to  argue that the AQNs may play the role of the  {\it triggers} initiating the  large scale  flares in the regions with large magnetic field.   To be more specific, we want to  argue that  the shock waves (which will be always generated as a result of high velocity  of the dark matter nuggets entering the solar atmosphere  with typical average $v_{\rm AQN}\sim 600 {\rm km/s}$ which is much greater\footnote{\label{free-fall}One should not confuse the typical velocities  of the DM particles $v_{\rm DM}\sim 10^{-3}c$ in galaxy with typical impact velocities of the DM particles entering the solar  atmosphere   as the free fall velocity ${v_{\rm AQN}}=\sqrt{\frac{2GM_{\odot}}{R_{\odot}}}\simeq 2\cdot 10^{-3}c\sim 600 ~{\rm km/s} $ is already very high for DM particles close to the surface of the Sun. Of course, there is an additional contribution related to the motion of the Sun around the galactic center $v_{\odot}\simeq 220~ {\rm km/s}$ and random velocity $v_{\rm random}\simeq 200~ {\rm km/s}$ of the DM particle distribution.} than the speed of sound $c_s$) can easily initiate the large flares as a result of the magnetic reconnection. 
 In section \ref{observations} we review  some observations, including x-ray data  supporting the basic framework. We also speculate that the sunquakes might be also related to the same AQNs initially entering from outer space and capable to reach   the photospheric layer.

\section{Axion Quark Nugget (AQN) dark matter model}\label{sec:QNDM}
 The AQN model in the title of this section stands for the axion quark nugget   model, see original work \cite{Zhitnitsky:2002qa}  and short overview 
\cite{Lawson:2013bya} with large number of references on the original results reflecting different aspects of the AQN model.   In comparison with many other similar proposals it has  two unique  features:\\
1. There is an  additional stabilization factor in the AQN  model provided    by the {\it axion domain walls}
  which are copiously produced during the QCD transition in early Universe;\\
  2. The AQNs  could be 
made of matter as well as {\it antimatter} in this framework as a result of separation of the baryon charges.
 
 The most  important astrophysical implication  of these new aspects   relevant for the present studies  (focusing on  different types of flares and  sources of the   EUV and  x-ray radiation  in the 
 solar chromosphere and corona)  
  is that quark nuggets made of  antimatter
 store a huge amount of energy which can be released when the anti-nuggets hit the Sun from outer space  and get annihilated.  This feature 
 of the AQN model is unique and is not shared by any other dark matter models because the dark matter in AQN model is made  of the same quarks and antiquarks of the standard model (SM) of particle physics. One should also remark here that the   annihilation events of the anti-nuggets with visible matter  may  produce a number of other observable effects in different  circumstances such as  rare events of annihilation of anti-nuggets with  visible matter     in the centre of galaxy, or in the    Earth atmosphere,  see some  references on the original computations in \cite{Lawson:2013bya}   and few comments at the end of this   section.

The basic idea of  the AQN  proposal can be explained   as follows: 
It is commonly  assumed that the Universe 
began in a symmetric state with zero global baryonic charge 
and later (through some baryon number violating process, the so-called baryogenesis) 
evolved into a state with a net positive baryon number. As an 
alternative to this scenario we advocate a model in which 
``baryogenesis'' is actually a charge separation process 
when  the global baryon number of the Universe remains 
zero. In this model the unobserved antibaryons come to comprise 
the dark matter in the form of dense nuggets of quarks and antiquarks in colour superconducting (CS) phase.  
  The formation of the  nuggets made of 
matter and antimatter occurs through the dynamics of shrinking axion domain walls, see
original papers \cite{Liang:2016tqc,Ge:2017ttc} with many technical  details. 

 The  nuggets, after they formed,  can be viewed as the  strongly interacting and macroscopically large objects with a  typical  nuclear density 
and with a typical size $R\sim (10^{-5}-10^{-4})$cm determined by the axion mass $m_a$ as these two parameters are linked, $R\sim m_a^{-1}$.
This  relation between the size of nugget $R$ and the axion mass $m_a$  is a result of the equilibration between the axion domain wall pressure and the Fermi pressure 
of  the dense quark matter  in CS phase. 
One can easily    estimate a typical  baryon charge $B$ of  such  macroscopically large objects as the typical density of matter in   CS phase  
is only few times the   nuclear density. 
However, it is important to emphasize that there are strong constraints on the    allowed window for the axion mass,  which can be represented as follows $10^{-6} {\rm eV}\leq m_a \leq 10^{-2} {\rm eV}$, see original papers \cite{axion,KSVZ,DFSZ} and   reviews \cite{vanBibber:2006rb,Asztalos:2006kz,Sikivie:2008,Raffelt:2006cw,Sikivie:2009fv,Rosenberg:2015kxa,Graham:2015ouw,Ringwald:2016yge}  on the theory of the axion and recent progress on axion search experiments. 

 This axion window corresponds to the range of the nugget's baryon charge $B$ which   largely overlaps  with all presently available and independent constraints on such kind of dark matter masses and baryon charges 
 \beq
 \label{B-range}
 10^{23}\leq |B|\leq 10^{28}, 
 \eeq
 see e.g. \cite{Jacobs:2014yca,Lawson:2013bya} for review\footnote{\label{B-constraint}The smallest nuggets with $B\sim (10^{23}-10^{24})$ naively  contradict to the constraints cited in \cite{Jacobs:2014yca}. However, the corresponding constraints are actually derived with the assumption that   nuggets with a definite mass (smaller than 55g) saturate the dark matter density. In contrast, we assume  that the peak  of the nugget's distribution corresponds to a larger value of mass, $\la B\ra \geq 10^{25}$, while the small nuggets represent a tiny portion of the total dark matter density. The same comment also applies to   the larger masses excluded by Apollo data as reviewed in  \cite{Lawson:2013bya}.
  Large nuggets with $B\sim 10^{28}$ may  exist, but represent  a small portion of the total dark matter density, and therefore, do not contradict the Apollo's constraints, see also some comments  on the baryon charge distribution  (\ref{distribution})  in   section \ref{AQN-nanoflares}.}.  
  The corresponding mass $\cal{M}$ of the nuggets  can be estimated as ${\cal{M}}\sim m_pB$, where $m_p$ is the proton mass, though more precise
  estimates relating the axion mass $m_a$, the baryon nuggets charge $B$ and the nugget's  mass $\cal{M}$ are also available \cite{Ge:2017idw}. 

 This  model is perfectly consistent with all known astrophysical, cosmological, satellite and ground based constraints within the parametrical range for 
 the mass $\cal{M}$ and the baryon charge $B$ mentioned   above (\ref{B-range}). It is also consistent with known constraints from the axion search experiments. Furthermore, there is a number of frequency bands where some excess of emission was observed, but not explained by conventional astrophysical sources. Our comment here is that this model may explain some portion, or even entire excess of the observed radiation in these frequency bands, see short review \cite{Lawson:2013bya} and additional references at the end of this section.

Another key element of this model is   the coherent axion field $\theta$ which is assumed to be non-zero during the QCD transition in early Universe.
       As a result of these $\cal CP$ violating processes the number of nuggets and anti-nuggets 
      being formed would be different. This difference is always of order of one effect   \cite{Liang:2016tqc,Ge:2017ttc} irrespectively to the parameters of the theory, the axion mass $m_a$ or the initial misalignment angle $\theta_0$. As a result of this disparity between nuggets and anti nuggets   a similar disparity would also emerge between visible quarks and antiquarks.  
       This  is precisely  the reason why the resulting visible and dark matter 
densities must be the same order of magnitude \cite{Liang:2016tqc,Ge:2017ttc}
\be
\label{Omega}
 \Omega_{\rm dark}\sim \Omega_{\rm visible}
\ee
as they are both proportional to the same fundamental $\Lambda_{\rm QCD} $ scale,  
and they both are originated at the same  QCD epoch.
  If these processes 
are not fundamentally related the two components 
$\Omega_{\rm dark}$ and $\Omega_{\rm visible}$  could easily 
exist at vastly different scales. 
 
  Unlike conventional dark matter candidates, such as WIMPs 
(Weakly interacting Massive Particles) the dark-matter/antimatter
nuggets are strongly interacting and macroscopically large objects,  as we already mentioned. 
However, they do not contradict any of the many known observational
constraints on dark matter or
antimatter    in the Universe due to the following  main reasons~\cite{Zhitnitsky:2006vt}:
  They carry  very large baryon charge 
$|B|  \gtrsim 10^{23}$, and so their number density is very small $\sim B^{-1}$.  
 As a result of this unique feature, their interaction  with visible matter is highly  inefficient, and 
therefore, the nuggets are perfectly qualify  as  DM  candidates. Furthermore, 
  the quark nuggets have  very  large binding energy due to the   large    gap $\Delta \sim 100$ MeV in  CS phases.  
Therefore, the baryon charge is so strongly bounded in the core of the nugget that  it  is not available to participate in big bang nucleosynthesis
(\textsc{bbn})  at $T \approx 1$~MeV, long after the nuggets had been formed.

  It should be noted that the galactic spectrum 
contains several excesses of diffuse emission the origin of which is unknown, the best 
known example being the strong galactic 511~keV line. If the nuggets have the  average  baryon 
number in the $\langle B\rangle \sim 10^{25}$ range they could offer a 
potential explanation for several of 
these diffuse components.  
\exclude{(including 511 keV line and accompanied   continuum of $\gamma$ rays in 100 keV and few  MeV ranges, 
as well as x-rays,  and radio frequency bands). }
It is important to emphasize that a comparison between   emissions with drastically different frequencies in such  computations 
 is possible because the rate of annihilation events (between visible matter and antimatter DM nuggets) is proportional to 
one and the same product    of the local visible and DM distributions at the annihilation site. 
The observed fluxes for different emissions thus depend through one and the same line-of-sight integral 
\be
\label{flux1}
\Phi \sim R^2\int d\Omega dl [n_{\rm visible}(l)\cdot n_{DM}(l)],
\ee
where $R\sim B^{1/3}$ is a typical size of the nugget which determines the effective cross section of interaction between DM and visible matter. As $n_{DM}\sim B^{-1}$ the effective interaction is strongly suppressed $\sim B^{-1/3}$. The parameter $\la B\ra\sim 10^{25}$  was fixed in this  proposal by assuming that this mechanism  saturates the observed  511 keV line   \cite{Oaknin:2004mn, Zhitnitsky:2006tu}, which resulted from annihilation of the electrons from visible matter and positrons from anti-nuggets.   Other emissions from different frequency bands  are expressed in terms of the same integral (\ref{flux1}), and therefore, the  relative  intensities  are unambiguously and completely determined by internal structure of the nuggets which is described by conventional nuclear physics and basic QED, see 
short overview  \cite{Lawson:2013bya} with references on specific computations of diffuse galactic radiation  in different frequency bands. 

   Finally we want to mention that the recent EDGES observation of a stronger than 
anticipated 21 cm absorption  \cite{Bowman:2018yin} can find an explanation within the AQN framework as recently advocated in 
\cite{Lawson:2018qkc}.  The basic idea is that the extra   thermal 
emission from AQN  dark matter at early times  produces the required intensity (without  adjusting of any parameters) to explain the recent EDGES observation.

  \section{The AQN annihilation events as nanoflares}\label{AQN-nanoflares} 
  We start our overview 
 of the basic results of ref \cite{Zhitnitsky:2017rop}  with subsection \ref{energetics} where we present  simple estimates 
    of the energetic budget    suggesting  that 
    the heating of the chromosphere and corona might be  due to the  annihilation events of the AQN with  the solar material. 
  As the next step, in subsection \ref{nanoflares}  we   highlight  the arguments of ref. \cite{Zhitnitsky:2017rop}  suggesting that the corresponding annihilation events
    heating the corona  can be  identified with the nanoflares conjectured  by Parker    long ago \cite{Parker}. Finally, in subsection \ref{nanoflares-development} we describe some recent observations of microflares and  flares (along with nanoflares) suggesting that all these phenomena might be in fact tightly connected and  originated from the same  AQNs in spite of the fact that these phenomena are characterized by drastically different energy scales: from $10^{20} \rm ~erg$  for nanoflares (much below the instrumental threshold being $\sim10^{24} \rm ~erg$) to  $10^{32} \rm ~erg$ for largest solar flares. 
    
    We elaborate on this proposal  in   the following section \ref{flares}   by demonstrating that the shock waves   will be always generated as a result of very high velocity of the AQNs entering the solar atmosphere  with $v_{\rm AQN}\simeq 600~ {\rm km/s}$, see footnote \ref{free-fall}. We suggest that these shock waves may serve as the triggers initiating the large solar flares if the AQNs enter the region of high magnetic field in which case the AQNs may activate the magnetic reconnection of {\it preexisted} magnetic fluxes.

  \subsection{Energetic budget due to the AQN annihilation events}\label{energetics}
  The impact parameter for capture of the nuggets by the Sun can be estimated as
  \be
  \label{capture}
  b_{\rm cap}\simeq R_{\odot}\sqrt{1+\gamma_{\odot}}, ~~~~ \gamma_{\odot}\equiv \frac{2GM_{\odot}}{R_{\odot}v^2},
  \ee
  where $v \simeq 10^{-3}c$ is a typical velocity of the nuggets far away from the Sun. One can easily see that  $\gamma_{\odot}\gg 1$ which implies that many AQNs which are not on head on collision trajectory, nevertheless will be eventually captured by the Sun.
   Nuggets in the solar atmosphere will  be decreasing their mass as result of annihilation, decreasing  their kinetic energy and velocity as result of ionization and radiation. 
  
    Assuming that $\rho_{\rm DM} \sim 0.3~ {\rm GeV cm^{-3}}$ and using the capture impact parameter (\ref{capture}), one can estimate 
  the total energy flux due to the complete annihilation of the nuggets,
   
  \be
  \label{total_power}
   L_{\odot ~  \rm (AQN)}\sim 4\pi b^2_{\rm cap}\cdot v\cdot \rho_{\rm DM}   
  \simeq 3\cdot 10^{30} \cdot \frac{\rm GeV}{\rm  s}\simeq 4.8 \cdot 10^{27} \cdot  \frac{\rm erg}{\rm  s}, 
  \ee
   where we substitute  constant $v\simeq 10^{-3}c$  to simplify numerical  analysis. 
   This estimate is very suggestive as it roughly coincides with the total EUV  energy output (\ref{estimate}) from corona which is hard to explain in terms of conventional astrophysical sources as highlighted in the Introduction. Precisely this ``accidental  numerical coincidence" was the main motivation   to put forward the idea \cite{Zhitnitsky:2017rop}
 that  (\ref{total_power}) represents a new source of energy feeding the EUV and x-ray radiation. 
       
    The main assumption made in   \cite{Zhitnitsky:2017rop} is that a finite portion of annihilation events have occurred before the anti-nuggets entered the dense regions of the Sun. This assumption has been recently supported by numerical Monte Carlo simulations \cite{Raza:2018gpb}
    which explicitly show that indeed, the dominant energy injection occurs in vicinity of the transition  region at the altitude $\sim 2000$ km.
   These annihilation events supply the energy source of the observed EUV and x-ray radiation from the corona and the choromosphere.
       The crucial observation made in   \cite{Zhitnitsky:2017rop} and confirmed  in \cite{Raza:2018gpb}
 is that  while  the total    energy due to the annihilation of the anti-nuggets 
   is indeed very small as it  represents $\sim 10^{-6}$ fraction of the solar luminosity according to (\ref{estimate}),  nevertheless the anti-nuggets  produce the EUV and x-ray spectrum  because the most of the annihilation events  occur in vicinity of the transition  region at the altitude $\sim 2000$ km 
   characterized by the temperature $T\sim 10^6$K.
       Such spectrum observed in corona and the chromosphere  is  hard to explain by any conventional astrophysical processes as
   mentioned  in the Introduction. 
   
   One should emphasize that the estimates (\ref{total_power})   for the total intensity  is not sensitive to the size distribution of the nuggets. This is because  the  estimate (\ref{total_power})   represents the total energy input due to the complete nugget's annihilation, while their total baryon charge is determined by the dark matter density  $\rho_{\rm DM} \sim 0.3~ {\rm GeV cm^{-3}}$
   surrounding the Sun. Explicit numerical analysis \cite{Raza:2018gpb} again confirms this claim.

    \subsection{Observation  of nanoflares as evidence for    anti-nuggets in Corona  }\label{nanoflares}
 
 In this subsection we highlight the arguments of ref. \cite{Zhitnitsky:2017rop} where the annihilation events of the anti-nuggets, which  generate  the energy (\ref{total_power}), are identified with the previously studied ``nanoflares", which belong to the burst-like solar activity.      
   The term ``nanoflare" has been introduced by Parker in 1983 \cite{Parker}. Later on this term has been used in series of papers by Benz and coauthors
  \cite{Benz-2000,Benz-2001, Kraev-2001,Benz-2002, Benz-2003}  and many others to advocate the idea that precisely these small ``micro-events" might be  responsible for the   heating of the   quiet solar corona.   It is not the goal of this work to review different aspects and different analyses related to  the nanoflares and the heating mechanisms. Instead, we just want to  mention few  papers on relatively  recent studies   \cite{Pauluhn:2006ut,Hannah:2007kw,Bingert:2012se,nanoflares} and reviews \cite{Klimchuk:2005nx,Klimchuk:2017} which support the basic  claim of early works that the nanoflares   play the  dominate role in heating of solar corona.
  However, some  disagreement still  remains between different groups on spectral properties of the nanoflares, see  some details below.

 We start this overview by providing the relation between the energy of the flares
 and the baryon charge $B$  of the AQNs.     Annihilation of a single baryon charge produces  the energy about 2 GeV which is convenient to express in terms of the conventional units as follows,
 \be
 \label{units}
 1~ {\rm GeV}=1.6 \cdot 10^{-10} {\rm J}=1.6\cdot 10^{-3} {\rm erg}.
 \ee 
In particular, this relation implies  that the current instrumental threshold of a nanoflare characterized by the energy  $\sim 10^{24}~{\rm erg}$ corresponds to the (anti) baryon charge of the nugget $B\approx 3\cdot 10^{26}$ which falls to  the window   (\ref{B-range}) of the allowed  baryon charges for the AQNs.  The nanoflares with sub-resolution energies (corresponding to   smaller values of $B$)  must be  present in the corona      to reproduce the measured radiation loss,  but are considered to be   the sub-resolution events, and cannot be resolved by the presently available instruments.

  Before we proceed with   the arguments  \cite{Zhitnitsky:2017rop} suggesting that the nanoflares can be interpreted as the annihilation events of the AQNs with the solar material, we would like to make few comments on the modern definitions of the nanoflares.  In most studies the term ``nanoflare" describes a generic event for any impulsive energy release on a small scale, without specifying its cause. In other words, in most studies the hydrodynamic consequences of impulsive heating (due to the nanoflares) have been used without discussing their nature, see review papers \cite{Klimchuk:2005nx,Klimchuk:2017}. The definition suggested in \cite{Benz-2003} is essentially equivalent to the definition adopted in  \cite{Klimchuk:2005nx,Klimchuk:2017} and refers to nanoflares as the ``micro-events" in quiet regions of the corona, to be contrasted with ``microflares" which are significantly larger in scale and observed in active   regions.   The term ``micro-events" refers to a short enhancement of coronal emission in the energy range of about $(10^{24}-10^{28})$ erg when  the lower limit   gives the instrumental threshold observing quiet  regions, while the upper limit refers to the smallest events observable in active regions.
  
  With these preliminary comments on definitions and units
  we want to highlight  below few important features which have been discussed in previous works  \cite{Benz-2000,Benz-2001, Kraev-2001,Benz-2002, Benz-2003}. We also want to  
 show  how these features are realized within the AQN framework  \cite{Zhitnitsky:2017rop}   advocating  the idea 
  that these nanoflares have the same properties  as   annihilation events  of   antinuggets in the corona. Therefore, the proposal of   \cite{Zhitnitsky:2017rop} is to identify them, i.e. 
  \be
  \label{identification}
  {\rm nanoflares}\equiv {\rm AQN~ annihilation~ events}.   
  \ee
  
  First of all, according to ref.\cite{Kraev-2001} to reproduce the measured  radiation loss, the observed range of nanoflares  needs to be extrapolated   from sub-resolution events with energy $3.7\cdot 10^{20}~{\rm erg}$ to the observed events  interpolating between   $(3.1\cdot 10^{24}  - 1.3\cdot 10^{26})~{\rm erg}$.    This energy window corresponds to the 
 (anti)baryon charge of the nugget $ 10^{23} \leq |B|\leq  4\cdot 10^{28}$  which largely  overlaps with allowed window   (\ref{B-range}) for AQNs reviewed  in section \ref{sec:QNDM}. We want to emphasize that this is a highly nontrivial consistency check for the proposal 
 (\ref{identification}) as the window (\ref{B-range}) comes from a number of different and independent  constraints extracted from 
 astrophysical, cosmological, satellite and ground based observations.  The window  (\ref{B-range}) is also consistent with known constraints from the axion search experiments within the AQN framework. Therefore, the overlap between these  two fundamentally different entities represents  a highly nontrivial consistency check of the proposal  (\ref{identification}).
 
 Our next comment goes as follows. According to ref.\cite{Benz-2003}    the nanoflares are {\it locally distributed very isotropically  in quiet  regions}, in contrast with micro-flares
 which are much more energetic and occur exclusively in the {\it local active areas}.  It  is perfectly consistent with our  identification (\ref{identification})  as the  anti-nugget annihilation events   should be present in all areas irrespectively to the activity of the Sun.  At the same time the micro- flares  are originated in the  active zones, and therefore cannot be isotropically  distributed. 
 
 One should comments here that {\it globally} there are few  obvious features  which may  influence of the global distribution  of the AQNs
 on the solar surface. First  feature is   related to the motion of the Sun with respect to the galactic center. This is because the velocity of the Sun  $|\vec{v}_{\odot}|\sim 220$ km/s is comparable with  typical randomly distributed  DM velocities $|\vec{v}_{\rm random}|\sim  200 ~{\rm km/s}$.
   Therefore, there will be {\it global}  preferential enhanced and suppressed regions on the solar surface which have drastically different number  
   of the AQN annihilation events. The corresponding global asymmetry $\Delta N$  due to the motion of the Sun  is determined by conventional expression for the flux of the DM particles in the framework of the Sun,
 \be
 \label{asymmetry}
\Delta N\sim  -\vec{v}_{\odot}\cdot \Delta\vec{S}    , ~~~~~~~~~ |\vec{v}_{\odot}|\sim 220 \frac{\rm km}{\rm s}
 \ee
where    $\Delta\vec{S}$  is element of area of the solar surface. The picture is in fact much more complicated because of the tilt of the Sun with respect to its velocity (about $30^o$) and because of the rotation of the Sun about its axis. 

The second feature which may also influence of the global distribution of the AQNs on the solar surface is the long range magnetic field with the correlation length of order $\sim R_{\odot}$. This is because the AQNs are the charged particles which are sensitive to external magnetic field. 
  Furthermore, the 
consequent evolution  of this injected  energy (due to the AQN annihilation events)  strongly varies  in different regions of the Sun 
due to the  intrinsic properties  of the solar atmosphere.

These global asymmetries are  likely to generate some local temporal  and  spatial inhomogeneities  of the AQNs on the solar surface. However, 
precise estimate of such asymmetries is well beyond of the score of the present work because
it requires specific Monte Carlo simulations  for each individual trajectory of the DM particle impacting the Sun when the typical time before impact is around  one month  \cite{Raza:2018gpb}. It is quite obvious that the changes of the initial  velocity vector of the AQN are order of one before impacting the Sun. This is precisely the reason why specific Monte Carlo simulations  for each individual trajectory are required to answer the question regarding the observed global asymmetries of the AQNs.

 Our next comment is related to the  observation  of the large Doppler shift  with a typical velocities (250-310) km/s, see Fig.5 in ref. \cite{Benz-2000}.  Furthermore, the observed line width in OV  of $\pm 140$ km/s far exceeds the thermal ion velocity which is around 11 km/s   \cite{Benz-2000}.
 These observed  features    can be  easily understood  within the AQN framework  and its   identification  proposal (\ref{identification}). Indeed, the typical initial velocities of the nuggets entering the solar atmosphere is about $ \sim  600 ~{\rm km/s}$, see footnote \ref{free-fall}. Therefore, it is perfectly consistent with observations of the  very large Doppler shifts and related broadenings of  the line widths.  Typical time-scales of  the nanoflare events, of order of $(10^1-10^2)$ sec. are also consistent with estimates  \cite{Zhitnitsky:2017rop}.

 One should also remark here that  the   energy output  observed by EIT on the SoHO satellite is of order of  $10\%$ of the total radiative output in the same region \cite{Benz-2002}. 
 The interpretation of this ``apparent deficiency" is   very straightforward within  our identification (\ref{identification}). Indeed, only a  small portion of the AQNs  are sufficiently large to produce the events with the energies above the instrumental threshold which can be recorded. Smaller events must also   occur and must   contribute to the total solar radiative output, but they are not recorded due to insufficient resolution of the current instruments.

Another comment goes as follows. The x-rays with the energies up to  $\sim 10$ keV have been observed in quiet regions, see  Fig. 9 in \cite{LZ:2003}. It is  hard to understand the nature  of such energetic photons from any conventional astrophysical sources. In our framework 
the x-rays are emitted by two different mechanisms in this framework. First, 
it is the direct consequence of the AQNs which  generate the shock waves 
with large Mach number $M\sim 10$ in quiet regions. The direct consequence of the shock wave is the increasing  of the local temperature $T\sim M^2$ along the nugget's jet-like shock front  as discussed in section \ref{shock}. The second mechanism is due to the direct annihilation of the antinuggets with the solar material as computed in \cite{Forbes:2006ba}, though in different context.

      Our last comment in this subsection is related to  nanoflare frequency distribution as a function of its energy.   The corresponding function  can be formally expressed as follows 
  \be
  \label{distribution}
   {dN} \sim B^{-\alpha_{\rm nano}}dB\sim W^{-\alpha_{\rm nano}} dW, ~~~~~ \rm for ~~~~ W\simeq  (4\cdot 10^{20}  -  10^{26})~{\rm erg}
  \ee
  where $dN$ is the number of the nanoflares  (including the sub-resolution events) per unit time with energy between $W$ and $W+dW$ which occur as a result of complete annihilation   of the anti-nuggets carrying the  baryon charges between  $B$ and $B+dB$.
 These two distributions are tightly linked  
    as these two entities are related to the same AQN objects according to (\ref{identification}).   
      The energy of the  events $W$ in this distribution can be always expressed   in terms of the baryon charges $B$ of the AQNs according to (\ref{units}).    
    
   The corresponding theoretical estimates of the distribution 
   $dN/dB$ are very hard to carry out  as explained in  \cite{Zhitnitsky:2017rop}. 
    Fortunately, on the observational (data analysis) side with the estimates  $dN/dW$ some  progress can be made, and in fact, has been  made.   
        In particular, the authors of ref. \cite{Pauluhn:2006ut} claim that the best fit to the data is achieved with $\alpha_{\rm nano}\simeq 2.5$, while numerous attempts to reproduce the data with $\alpha_{\rm nano} < 2$ were unsuccessful. This is consistent with previous analysis 
 \cite{Benz-2002} with $\alpha_{\rm nano}\simeq 2.3$.
It should be contrasted with another analysis \cite{Bingert:2012se} which suggests that $\alpha_{\rm nano}\simeq 1.2$ for events below $W \leq 10^{24}$ erg,   and $\alpha_{\rm nano}\simeq 2.5$ for events above $W \geq 10^{24}$ erg.  Analysis \cite{Bingert:2012se}  also suggests that the change of the scaling (the position of the knee) occurs at energies close to $\la W\ra \simeq 10^{24}~ {\rm erg}$, which roughly coincides with the maximum of the energy distribution, see Fig.7 in   \cite{Bingert:2012se}. 

We conclude this subsection with the following comment. While  there is general agreement that the nanoflares (including the sub-resolution events) are responsible for the heating of the corona, there is some  disagreement between different groups  on spectral properties $dN/dW$ of the flares expressed in terms of power-law index $\alpha_{\rm nano}$ as defined by  (\ref{distribution}).

 \subsection{From nanoflares $\Rightarrow$ microflares $\Rightarrow$ large solar flares}\label{nanoflares-development}
 The question we want to address in this subsection can be formulated as follows.
 The nanoflares, as discussed in the previous subsection, can be identified with  the AQNs according to (\ref{identification}).
 Furthermore, the nanoflare events (or what is the same the AQN annihilation events) is the main source of the heating corona and 
 as the consequence, the source of the observed EUV radiation (\ref{total_power}) in agreement with observations  (\ref{estimate}). 
 At the same time, there is a strong argument presented in the Introduction which suggests  that the EUV radiation is strongly correlated with large M, X -flares, see Fig. \ref{EUV} and ref.\cite{Zioutas} with details. Therefore, one should expect that these two phenomena (nanoflares versus M,X-flares) characterized by drastically different scales, nevertheless must be originated from the same physics. 
 
 The  natural question occurs: how it could be ever  possible that the M, X -flare events which are characterized by  energy scale $W \simeq (10^{26}- 10^{32})$ erg   could be related to   relatively small objects with energies  
  (\ref{distribution}) which   describe the  nanoflares? Such potential  relation becomes even more suspicious    if 
  one recalls that the 
  nanoflares are distributed very uniformly in quiet regions, as discussed in previous section. It should be   contrasted with microflares and large flares which are much more energetic and occur exclusively in active areas. 
 
 Our proposed answer on this question will be given  at the end of the section. Before we formulate our proposed answer   we want to mention that 
 the idea that small nanoflares and large solar flares might be  originated from the same physics is not very new, and  has been discussed   previously in  the literature. The basic argument is based on analysis of 
   the large flare frequency distribution, similar to   (\ref{distribution}), and is defined as   follows 
 \be
  \label{distribution1}
   {dN} \sim   W^{-\alpha} dW, ~~~~~~~~~ \rm for ~~~~ W\simeq  ( 10^{26}  -  10^{32})~{\rm erg},
  \ee
  with the only difference in comparison with  (\ref{distribution}) is that the energy  $W$  covers  the large flare  region. 
  We want to a mention two different analysis: ref.\cite{Hannah:2007kw} which includes RHESSI data with the energies range from $10^{26}~ {\rm erg}$
to $10^{30}~ {\rm erg}$, and  ref. \cite{Shibata:2016} where even super flares with energies $10^{34}~ {\rm erg}$
to $10^{35}~ {\rm erg}$ have been considered. 
 
  It has been noted in \cite{Hannah:2007kw} that it is conceivable 
that the distribution of all flares follows a single power-law with $\alpha\simeq \alpha_{\rm nano}\simeq 2$, which might suggest a common origin for all flares, see Fig. 18 in \cite{Hannah:2007kw}.  Of course, the comparison between different components of the energy distribution is a  highly  nontrivial procedure as it includes comparison of the data produced by the different instruments with specific instrumental effects. Furthermore,   different components of the energy distribution covers different phases of  the solar cycle. Analysis \cite{Shibata:2016} also suggests that all flares (including the superflares) can be described by a single power with $\alpha\simeq \alpha_{\rm nano}\simeq 1.8$, see Fig. 18 in ref.
\cite{Shibata:2016}. 

   From the AQN dark matter model perspectives   the nanoflares and large flares 
 are considered to be very different types of events. The nanoflares are identified with AQN according to (\ref{identification}) and must be distributed uniformly through the solar surface, which is precisely what has been observed as reviewed in previous subsection. 
 All larger flares   distributed very non-uniformly on the solar surface and localized in the active regions (sunspots) which are characterized by a  strong magnetic field representing the source of the large  flares. 
 
 Precisely this   distinct feature  in spatial distribution  constituents the answer on the question formulated at the beginning  of this subsection:    the   antinuggets  (distributed uniformly)  play the role of the {\it triggers}   activating   the magnetic reconnection of {\it preexisted}  magnetic fluxes in {\it active regions}.  This relation represents the link between the nanoflares and large flares. This  proposed link  is obviously very different from the arguments  reviewed above and based on similarities between the exponents  $\alpha$ and $\alpha_{\rm nano}$.
 
 Within our scenario the energy of the flares  is generated by the  preexisted magnetic field occupying very large area in active region, while relatively small amount of energy associated with initial AQNs (nanoflares) play a minor role in the total  energy released during a large flare. 
 Within this framework all large flares (including microflares) follow a single power-law with $\alpha\simeq 2$ while nanoflares represent very different type of events which, in general, are characterized (from the AQN perspective)   by a different exponent $\alpha_{\rm nano}$\footnote{Different analysis leading to different $\alpha_{\rm nano}$ as discussed after eq. (\ref{distribution}), do not  affect 
 the results of the present work. In particular, there is no contradiction of our proposal leading to estimates $\alpha\simeq 2$ presented in section  \ref{scaling} with the analysis of  \cite{Bingert:2012se}   where  $\alpha_{\rm nano}\simeq 1.2$ for events below $W \leq 10^{24}$ erg.   As we mentioned above, the nanoflares  and flares are different  
 entities in this framework. The nanoflares are uniformly distributed through the solar surface, and they are powered by the annihilation energy of the AQNs. It should be contrasted with large flares powered by the   magnetic energy  localized  in active regions.}. 
 One should also comment here that the same tendency with approximately the same power law $\alpha\simeq 2$ holds for 
 superflares in solar type stars, see Fig.18, references and discussions in ref.\cite{Shibata:2016}. This similarity suggests that the superflares 
 are originated from the same physics. In fact,  our  arguments supporting $\alpha\simeq 2$ in section  \ref{scaling}  have pure geometrical  nature and can be equally applied to  solar flares as well as to superflares occurring in the solar-type stars.

 We elaborate on this proposal in next section by demonstrating that the AQNs entering the corona from outer space  will generate the {\it shock waves}, playing the role of the triggers, as the velocity of the nuggets $v_{\rm AQN}\sim 10^3 {\rm km/s}$ is well above the speed of sound, $v_{\rm AQN}/c_{\rm sound}\gg 1$.   The only comment we would like to make here is that this proposal is consistent with the observations that  the typical time scale of the nanoflares is $(10-100)$ sec. while the large flares last much longer. Furthermore, the typical length scales  for these phenomena are also drastically different: the nanoflares are characterized by the scale of order $10^3$ km, while a typical characteristic for large flares is  around  $(10^4-10^5)$ km. 
 These drastic differences  in time and length  scales support the idea that AQNs  serve as the triggers initiating the larger flares. In this case    the time scales of large flares are related to the physics of the magnetic reconnection (measured in hours), in contrast with  the   annihilation rate   of the AQN (measured in seconds), which assumes   a typical scale for nanoflares irrespectively to the region where annihilation occurs: whether it is a  active or quiet region.

To conclude this section one should emphasize that the corresponding  asymmetries  mentioned in subsection  \ref{nanoflares}   are  expected to be very tiny on the level of few percents such that  the  basic AQN distribution remains to be very isotropic. It should  be contrasted with highly asymmetric distribution of the  solar flares discussed in subsection \ref{nanoflares-development}.  The large solar flares  are strongly correlated with sunspot areas according to this proposal formulated in next section \ref{flares}. The sunspots in active regions are obviously distributed with high level irregularities as the corresponding physics is entirely determined by internal solar dynamics. Therefore, the solar flares are also distributed very 
non-uniformly. Needless to say that all known correlations such that the butterfly diagram, frequency of   flares during the solar cycles, the longitude dependence, etc
are entirely determined by the  internal solar dynamics and cannot be affected by the  AQNs which  are drastically  less energetic objects, and  which  merely play the role of the triggers of large solar flares as will be discussed below.

 \section{AQNs as the triggers initiating the  solar flares}\label{flares} 
 We start   in subsection \ref{SP}  with a short overview   of old Sweet-Parker's theory   on the magnetic reconnection, its development,  its results, its problems and  difficulties. In subsection \ref{shock}  we present the  arguments suggesting  that the AQNs entering the solar corona will inevitably  generate the shock waves as a result of high velocity of the nuggets $v\sim 10^{-3}c$. 
 One should note that the shock waves will be produced by both kind of species: nuggets and antinuggets.
 Therefore, in the rest of the paper we do not distinguish different species, in contrast with our discussions of the corona heating proposal \cite{Zhitnitsky:2017rop,Raza:2018gpb}  reviewed in previous section \ref{AQN-nanoflares} when the annihilation energy 
 (due to the antinuggets) plays the key role in the arguments. 
 
 In the  next subsection \ref{m_reconnection}   we argue  that precisely these shock waves may serve as the  triggers which are capable to initiate (ignite)  the magnetic reconnections and generate the large flares. Finally, in subsection   \ref{scaling} we argue that the observed scaling (\ref{distribution1}) with $\alpha\simeq 2$  can be interpreted in pure geometrical way.  
 
 \exclude{The section \ref {flares} will prepare us for following section  \ref{observations} where we confront  the immediate and direct  consequences  of this proposal with observations. We find that all qualitative  features  which are  basic   consequences  of this  framework are consistent with large number of   known observations collected by  different instruments 
 during many years of studies of the solar flares.  
 }
 \subsection{Magnetic reconnection and  Sweet-Parker theory}\label{SP}
 We start by introducing the most important parameters of the problem
 \be
 \label{definitions}
 S=\frac{Lv_A}{\chi_m}, ~~~~~ v_A=\frac{\cal{B}}{\sqrt{4\pi\rho}}, ~~~~~ \chi_m=\frac{c^2}{4\pi\sigma}, 
 \ee
where  $S$ is the so-called the Lundquist number,   $L$ is the typical size of the problem,   $v_A$ is Alfv\'{e}n speed, $\rho$ is the plasma's mass density,  $\chi_m$ is the magnetic diffusivity, and finally  $\sigma$ is the electrical conductivity of the plasma. 
The most important parameter for our future estimates is the dimensionless parameter $S$ which assumes the following values for typical coronal conditions, $S\sim (10^{12}-10^{14})$. 

Original idea on magnetic reconnection was formulated   by Sweet \cite{Sweet}  and Parker \cite{Parker1} sixty years ago.
Using simple dimensional arguments, Sweet and Parker (SP) have shown that the reconnection time $\tau_{\rm rec}$ is  
quite slow and expressed in terms of the original parameters of the system as follows
\be
\label{eq:SP}
\frac{\tau_A}{\tau_{\rm rec}}\sim \frac{1}{\sqrt{S}}, ~~~~ \tau_A\equiv\frac{L}{v_A}, ~~~~ \frac{u_{\rm in}}{u_{\rm out}}\sim  \frac{1}{\sqrt{S}}, ~~~~ \frac{l}{L}\sim \frac{1}{\sqrt{S}}, ~~~~~\tau_{\rm rec}\sim \frac{L}{u_{\rm in }}
\ee
 where $u_{\rm in}$ is the  velocity of reconnection between oppositely directed fluxes of thickness $l$, and  $u_{\rm out}\sim v_A$ is normally assumed to be of   order of  the Alfv\'{e}n velocity. The scaling relations (\ref{eq:SP}) predicted by SP theory are  obviously insufficient to explain the reconnection rates observed in corona due to the very large numerical values of $S\sim (10^{12}-10^{14})$.
 
 The next step to speed up the reconnection rate has been undertaken in \cite{Petschek} with some important amendments in \cite{Kulsrud} where it was  argued that the reconnection rate could be much faster than the  original formula (\ref{eq:SP}) suggests.
 However, some subtleties remained in the proposal \cite{Petschek,Kulsrud}. Furthermore, the numerical simulations reproduce conventional scaling formula    (\ref{eq:SP}) at least for moderately large $S\lesssim 10^4$. In the last 10-15 years 
  large number of new ideas have been pushed forward. It includes, but not limited to such processes as plasmoid- induced reconnection,  fractal reconnection, to name just a  few. 
 
 It is not the goal of the present work to analyze  the assumptions, justifications, and the problems related to the old proposals 
   \cite{Sweet,Parker1,Petschek,Kulsrud} and new ideas,  and we refer to 
 the recent review papers \cite{Shibata:2016,Loureiro} for recent developments and relevant comments on these matters.  The only comment we would like to make here is that the new element   (we are advocating in the present work) is the presence of an additional ingredient     in the problem which was not a part in the previous studies. This new ingredient of the problem   is the AQNs which enter the system from outer space and generate the shock waves in corona as we argue in next subsection \ref{shock}. Our proposal is  that precisely these shock waves will serve as the triggers   initiating  the magnetic reconnections
 \cite{Sweet,Parker1,Petschek,Kulsrud,Shibata:2016,Loureiro}
  which eventually lead to large solar flares.

 \subsection{AQNs and  shock waves in plasma}\label{shock}
 In this subsection we argue that the AQNs entering the solar corona will inevitably generate the shock waves as a result of high initial velocity of the nuggets $v_{AQN}\sim 600~ {\rm km/s}$ on the solar surface, see footnote \ref{free-fall}. To simplify the arguments and notations in our presentation   below   we   do not include the magnetic field into consideration at this point\footnote{\label{neglect}Neglecting the magnetic field  is sufficiently good approximation  in quiet regions when  the magnetic energy does not dominate the dynamics. In this case
   the additional energy is  generated exclusively as  a result of  annihilations   of the AQNs with the solar material, while the energy of the magnetic field plays a minor role.   The corresponding annihilation events of the AQNs   are identified with nanoflares (\ref{identification}), and  the  extra energy  due to the annihilation   is released as the EUV and x ray radiation as reviewed  in section \ref{AQN-nanoflares}.}. 
 The corresponding generalization with inclusion of  a strong  magnetic field (relevant for analysis of the AQNs in the active regions)   will be presented in the following section \ref{m_reconnection}. 
 
 In the present  subsection we assume that the magnetic field $B\sim 1$~G which is a typical magnitude in nonactive region, far away from sunspots.   In this case the magnetic pressure much smaller than the plasma pressure.   Furthermore, is Alfv\'{e}n speed $v_A$ speed,  to be estimated below,  is also very small, much smaller than speed of the nuggets, $v_A\ll v_{AQN}$. In these circumstances 
 one can approximate the plasma ignoring the magnetic field contribution, see footnote \ref{neglect}.

 We start our analysis by estimating   the speed of sound $c_s$ in corona at $T\simeq 10^6 K$, 
 \be
 \label{sound}
 \left(\frac{c_s}{c}\right)^2\simeq \frac{3\gamma T}{m_p},  ~~~ c_s\simeq 7\cdot 10^{-4} c\cdot \sqrt{ \frac{T}{10^6~ {\rm K}}}, ~~~
 c_s\simeq 2\cdot 10^7 \sqrt{ \frac{T}{10^6~ {\rm K}}}\cdot  \left(\frac{\rm cm}{\rm s} \right), 
 \ee
 where $\gamma=5/3$ is a specific heat ratio, and we approximate the mass density $\rho$ of plasma by the proton's number density density $n$ as follows $\rho\simeq n m_p$. The crucial observation here is that the Mach number $M$ is always   larger than one for a typical dark matter velocities:
 \be
 \label{Mach}
 M\equiv \frac{v_{AQN}}{c_s} \geq 5  \sqrt{ \frac{10^6~ {\rm K}}{T}} \gg 1.
 \ee
 As a result, the shock waves will be  inevitably generated when the AQNs enter the solar corona. 
 
 In the limit when the thickness of the shock wave can be ignored  the corresponding formulae for  the discontinuities of the  pressure $p$, temperature $T$, and the density $\rho$ are well known and given by, see e.g. \cite{Landau}
 \be
 \label{eq:shock}
 \frac{\rho_2}{\rho_1}&=&\frac{(\gamma+1)M^2}{(\gamma-1)M^2+2},  \\  \frac{p_2}{p_1}&=&\frac{2\gamma M^2}{(\gamma+1)}-\frac{\gamma-1}{\gamma+1}, \nonumber \\ \frac{T_2}{T_1}&=&\left(\frac{p_2}{p_1}\right)\cdot \frac{(\gamma+1)p_1+(\gamma-1)p_2}{(\gamma-1)p_1+(\gamma+1)p_2}  \nonumber.
 \ee   
 For our qualitative analysis which follows we assume that the Mach number $M\gg 1$, in which case the relations 
 (\ref{eq:shock}) are greatly simplified  and assume  the form
 \be
 \label{shock1}
 \frac{\rho_2}{\rho_1}\simeq \frac{(\gamma+1) }{(\gamma-1)}, ~~~~~  \frac{p_2}{p_1}\simeq M^2\cdot \frac{2\gamma }{(\gamma+1)}, ~~~~~ \frac{T_2}{T_1}\simeq M^2\cdot \frac{2\gamma(\gamma-1) }{(\gamma+1)^2}.
 \ee   
 The relations (\ref{shock1}) imply that the discontinuities in temperature $T$ and  pressure $p$ could be numerically enormously large   as they are proportional to the Mach number $M^2$ which itself  could assume  very  large number  in the given circumstances  according to (\ref{Mach}).
 
 In such a regime (with   large   Mach numbers $M\gg 1$) the conventional hydro-based computations of the thickness width $\delta$ and the absorption coefficient may not be justified and true microscopical computations might be required in this  case \cite{Landau}. 
 Indeed, one can estimate the thermal velocities $v_T$ of the particles in the plasma, kinematic viscosity $\nu$, thickness of the of shock wave $\delta$ and the sound absorption coefficient $a$ at $M\gg 1$  to convince yourself that all the relevant parameters are expressed in terms of the mean free path $l_0$:
 \be
 \label{parameters}
 v_T\sim \sqrt{\frac{T}{m_p}}\sim c_s, ~~~ \nu\sim l_0 c_s, ~~~ a\sim \frac{l_0}{c_s^2}, ~~~~ \delta\sim l_0.
 \ee
  The estimates (\ref{parameters}) unambiguously imply that the hydro-based computations cannot be used to study strong shock waves with $M\gg 1$ because the basic assumption that the  plasma can be considered as a continuous media (when   $l_0\rightarrow 0$ is assumed to be a smooth limit)  cannot be justified. 
 
 \exclude{
 The mean free path $l_0$ in corona can be estimated as follows. By definition, $l_0^{-1}\sim \sigma n$, where $n$ is the number density in plasma, and $\sigma\sim \alpha^2/q^2$ is the  Coulomb cross section for a typical momentum transfer $q$. The momentum transfer $q$ can be estimated as a typical temperature $T$ of the  plasma, i.e. $q\sim T$.  Collecting all factors together we arrive to the following estimate for $l_0$:
 \be
 \label{l_0}
l_0^{-1}\sim \sigma n\sim \frac{n\alpha^2}{T^2}, ~~~ 
 \ee  
  Our numerical estimates (\ref{l_0}) suggest that the thickness width $\delta$ of the shock wave can be estimated as follows 
  \be
  \label{delta}
  \delta\sim l_0\sim 10^2 {\rm cm}\cdot \left(\frac{10^{10}~ {\rm cm^{-3}}}{n}\right)\cdot \left(\frac{T}{10^6~K}\right)^2,
  \ee 
  which we think is quite  reasonable  number. 
  }
  
 What is the role of the shock waves which will inevitably form when the AQNs enter the solar atmosphere?
  The total energy 
  due to the annihilation events of the AQNs (identified with nanoflares  according to (\ref{identification}))  is fixed and determined by the dark matter density according to the  estimate (\ref{total_power})  which  agrees  with observations (\ref{estimate}).   As we discussed in \cite{Zhitnitsky:2017rop} and reviewed in section \ref{AQN-nanoflares} this energy is sufficient to heat the corona to $T\simeq 10^6$K. There are many mechanisms which are capable to  transfer energy from AQNs to plasma. The shock waves and the turbulent boundary layer which always accompanies  a body moving with $M>1 $  eventually must play a key role in this energy transfer from AQNs (nanoflares)    to the plasma\footnote{The corresponding  hard questions on energy transfer are well beyond the scope of the present work.   
  The only comment we would like to make here is that the corresponding estimates are  likely to require a truly microscopical computations as mentioned above.}.  
  
   Apparently, a structure which resembles very much 
  a jet (which is a typical shape  of a shock wave) 
   has been observed in the chromosphere \cite{Shibata:2007}, see also review \cite{Shibata:2016}. We'll  discuss the observational consequences of our  proposal in more details in subsection \ref{observations}. Now we want to make a short comment that the observed jets, coined as  ``chromosphere anemone jets"  in 
    \cite{Shibata:2007} are  few thousand kilometres long and few hundred kilometres wide.

   From the AQN perspective such a cone-like shape  indeed represents a  typical  morphology  for a   body moving with $M>1$ which  generates a shock wave. The observed tiny  jets are also characterized by a typical for nanoflares   energy (\ref{distribution}) which can be interpreted as an additional 
    indirect  support of  our identification (\ref{identification}) of AQNs generating the shock waves with nanoflares conjectured long ago \cite{Parker}. Though, one should be very careful with identification of any observed jets with the jets produced by the AQNs because 
    the dominant portion of the AQNs is characterized by the baryon charges in the window (\ref{B-range})  which is well  below the instrumental threshold and, therefore,  could not be directly observed.

 \subsection{Magnetic reconnection ignited by the shock waves}\label{m_reconnection} 
 The question we address in this section can be formulated as follows: what happens if the AQNs generating the shock waves
 with $M\gg 1$ as discussed above, will enter the active region with sufficiently large magnetic field?  In this case the dynamics of the magnetic field cannot be ignored and must be included into the consideration.  There are few  important parameters which control  the dynamics of the  system: in addition  to  already defined parameters (\ref{definitions})  it is convenient to introduce another dimensionless  parameter $\beta$ which determines the importance of the magnetic pressure,  
 \be
 \label{beta}
 \beta\equiv \frac{8\pi  p}{{\cal{B}}^2} \sim 0.5\cdot 10^{-1}\left(\frac{n}{10^{10} ~{\rm cm^{-3}}}\right)\cdot\left(\frac{T}{10^6 K}\right)\cdot\left(\frac{100~ G}{{\cal{B}}}\right)^2, 
 \ee
 where for numerical estimates we use typical parameters for the active regions in corona when $\beta\ll 1$.  
 Another important parameter   is Alfv\'{e}n speed $v_A$ which assumes the following numerical value in this environment
 \be
 \label{alfven}
 \frac{v_A}{c}=\frac{{\cal{B}}}{c\sqrt{4\pi\rho}} \sim \left(\frac{{\cal{B}}}{100~ G}\right)\cdot \sqrt{ \frac{10^{10} ~{\rm cm^{-3}}}{n} } \sim 2\cdot 10^{-3}, ~~~~ v_A\simeq 600  ~\frac{\rm km}{\rm s}. 
 \ee 
 For these parameters the numerical value for  $v_A$ is three times larger than   the speed of sound $c_s$ according to (\ref{sound}).  
 One can introduce the so-called Alfv\'{e}n Mach number $M_A$ which plays a role similar to the Mach number and  defined as follows
 \be
 \label{M_A}
 M_A\equiv \frac{v_{\rm AQN}}{v_A}.
 \ee
  In general $M_A\sim 1$ in active regions because typical $v_{\rm AQN}\simeq  600~ {\rm km/s}$  assumes the same order of magnitude as $v_A$, see footnote \ref{free-fall}.
   The $M_A$ could be slightly grater than one, or it could be slightly lower than unity, depending on magnitude of the magnetic field in a specific active region.  To simplify our qualitative analysis in what follows we assume that some finite fraction of the nuggets will have $M_A > 1$, similar to our assumption that $M\gg 1$. This assumption implies that we consider the AQNs with sufficiently high velocities toward the Sun. In this case one should expect that  the fast shock may develop which may trigger the large flare\footnote{\label{velocity}This is not very restrictive constraint as   
  the free fall velocity $   {v_{\rm AQN}}=\sqrt{\frac{2GM_{\odot}}{R_{\odot}}}\simeq 2\cdot 10^{-3}c\sim 600 ~{\rm km/s} $ is already very high. In addition,    the  velocity of the Sun  $|\vec{v}_{\odot}|\sim 220$ km/s is also large. Furthermore,  the  typical random DM velocity in the galactic halo $|\vec{v}_{\rm random}|\sim 200$ km/s. These estimates suggest that some finite  fraction of AQNs will have $M_A>1$, 
  while another   fraction of AQNs will have      $M_A<1$.}.

 With these preliminary comments we are in position to formulate the key question of this section: what happens  when the AQN enters the region with $\beta\ll 1$ and generates the shock wave with $M\gg 1, M_A>1$?
 The proposed answer is that the discontinuities in  plasma pressure,   magnetic  pressure, and the temperature  due to the shock wave will compress the region where the magnetic fields have opposite  directions and where  the magnetic reconnection starts. Precisely this is the region  where    the current sheet is generated, which eventually transforms the energy of the magnetic field into the radiation.  
 If it happens it would answer the main question formulated in the Introduction: how it is possible that the two drastically different phenomena 
 (the EUV radiation and the large flares) are both correlated with positions of the planets.  
  
 Before we continue with our numerical estimates we must say that the idea   that the shock waves may drastically increase the rate of magnetic reconnection is not new, and have been discussed previously in the literature \cite{Tanuma}, though in quite different context: it was applied to interstellar medium in the presence of the supernova shock. The new element which is advocated in the present work is that the small shock waves  resulting from  entering the AQNs are widespread and generic events in solar corona within  AQN  dark matter scenario. 
 Precisely these generic annihilation events being  identified with nanoflares  (\ref{identification})   are  responsible for the corona heating and the EUV emission. In addition, 
 as a result of this generality the  large flares (which are generated as a result of the magnetic reconnections) 
 might be also sparked by the same sufficiently fast AQNs. 
The most important part of this work that these  large flare  events   must be correlated with dark matter flux according to this AQN framework.  This relation explains how the intensity for the EUV emission and the frequency 
 of the large flares are interlinked and correlated with positions of the planets. This is precisely the correlation which has    been analyzed   in \cite{Zioutas}, and which was the main motivation for the present studies.  
 
 The key element in our arguments suggesting that the fast shock  wave may drastically modify  the rate of magnetic reconnection
 is based on observation that the pressure,  the temperature and the magnetic pressure  may   experience some dramatic changes when the shock wave approaches the reconnection region. To be more precise, the  formulae (\ref{eq:shock}) get modified in the presence  of  the magnetic field  as follows \cite{Landau}
 \be
 \label{delta-pressure}
  \frac{p^*_2}{p^*_1}&=&\frac{2\gamma M^2}{(\gamma+1)}-\frac{\gamma-1}{\gamma+1}, ~~~~~ {\rm where} ~~~~  p_i^*\equiv p_i+\frac{{\cal{B}}_i^2}{8\pi}
  \nonumber \\ \frac{T_2}{T_1}&=&\left(\frac{p^*_2}{p^*_1}\right)\cdot \frac{(\gamma+1)p^*_1+(\gamma-1)p^*_2}{(\gamma-1)p^*_1+(\gamma+1)p^*_2}~~,  
 \ee
 where subscript $p_1$ corresponds to the unperturbed  system without shock wave. 
 \exclude{ A result of the magnetic pressure 
 during the short period of time of passage of  the shock wave   through the  (future) reconnection region.
 The main observation here is that assuming that the shock is fast it may trigger the fast reconnection.

 the extra pressure strongly affects (locally) the parameter $\beta$, 
 \be
 \label{delta-beta}
 \Delta\beta\equiv \frac{8\pi  \Delta p}{{\cal{B}}^2} \sim 0.5\cdot 10^{-1}\cdot \left(\frac{2\gamma M^2}{\gamma+1}\right)\cdot \left(\frac{n}{10^{10} ~{\rm cm^{-3}}}\right)\cdot\left(\frac{T}{10^6 K}\right)\cdot\left(\frac{100~ G}{{\cal{B}}}\right)^2.
 \ee
 Similar modifications also occur for the density $\Delta \rho$ and the temperature $\Delta T$ according to eqs. (\ref{shock1}). 
 The crucial  observation here is that the effective parameter $\Delta\beta$ given by (\ref{delta-beta}) can become numerically very large
 in contrast with unperturbed case (\ref{beta}) when this parameter is typically very small.   Indeed, for $M^2\gtrsim 10$ which can be easily achieved according to estimate (\ref{Mach}), the parameter $\Delta\beta >1$ becomes larger than one. In fact it could become enormously large.
 }
  
   If the Mach numbers are large:  $M\gg 1,  M_A>1$, which we assume to be the case,   
    the magnetic flux  can be easily pushed into the direction  to the current sheet  region  where the reconnection starts. Precisely this scenario  (when $\Delta T/T\gg 1$ and $\Delta p/p\gg1 $ in spatially very small region for a very short period of time) represents a proposed mechanism for the  reconnection triggered and initiated by the shock wave.
    This happens if   the relative  AQN impact velocity   on the solar surface    is  sufficiently high.
     
 Now we want to estimate the size and the energy scales associated with  such events. We consider separately two different stages.  First, we estimate  the scales related to the initial phase of the evolution when the AQNs produce the shock waves,  but the magnetic reconnection has not started yet.  The estimation for the second phase assumes that the magnetic reconnection, leading to a large solar flare,   is already  fully developed.  
 
 In the first, initial stage of the evolution, the magnetic reconnection has not started yet, and entire energy is related to the shock wave, which itself forms as a result of AQN entering the solar atmosphere from outer space. In this case a typical time scale when AQN completely annihilates its baryon charge is   of order of $\tau\sim 10~ {\rm sec}$, see \cite{Zhitnitsky:2017rop,Raza:2018gpb}. A typical length scale is determined by the initial velocity   of the AQN which is of order $v_{\rm AQN}\sim  (600 -700){\rm km/s}$ such that $L\sim    v_{\rm AQN}\cdot \tau \sim 5\cdot 10^3~ {\rm km}$. At the same time, a typical radius $R$ of the cone formed by the shock wave is determined by the speed of sound $c_s$, such that $R\sim M^{-1} L$, where Mach number $M$ is estimated in (\ref{Mach}). For numerical estimates below we take $M\simeq 10$. The affected volume of the cone due to the shock wave is estimated as $V\simeq  ( \pi R^2L)/3\sim 10^{-2}L^3$. 
 We  summarize the parameters of the initial stage as follows
 \be
 \label{initial}
 \tau\sim 10~ {\rm sec}, ~~ L\sim  5\cdot 10^3~ {\rm km}, ~~ R \sim 10^{-1}L, ~~ V\sim \frac{\pi R^2L}{3}\sim 10^{-2}L^3.
 \ee
 We are now in position to estimate the typical energetic characteristics of the system during this {\it initial} stage. 
 The key element is the observation that the temperature $T$ experiences a large discontinuity resulting form the shock according to (\ref{delta-pressure}). Therefore, we estimate a typical temperature   as follows
 \be
 \label{initial1}
 \frac{T_2}{T_1}\sim M^2 \sim 10^2,  ~~~\Delta T\equiv (T_2-T_1)\sim M^2 T_1\sim  10^8~ {\rm K}
 \ee
  where $T_1\sim 10^6 K$ corresponds to unperturbed temperature before the shock passage through the area. 
  \exclude{
  One can estimate the   energy disturbance related to the shock wave $ E_{\rm shock}$ during the initial stage  of evolution 
  as follows  
  \be
 \label{initial2}
 E_{\rm shock}&\sim& n\Delta T V\sim 10^{28}{\rm erg}\cdot
  \left(\frac{M  }{10}\right)^2\cdot \left(\frac{n}{10^{10} ~{\rm cm^{-3}}}\right)\cdot\left(\frac{T_1}{10^6 K}\right)\cdot \left(\frac{V_{\rm shock}}{10^9{\rm km^3}}\right), 
 \ee
One should emphasize that $ E_{\rm shock}$ should not be interpreted as  the extra energy   generated by the AQN.
In fact,  $E_{\rm shock}$ is many orders of magnitude larger that the energy associated with AQN itself, which carries a  typical  nanoflare energy reviewed in section \ref{nanoflares}. 
 Instead, $E_{\rm shock}$  should be interpreted as a result of the redistribution of the temperature, pressure and the density between different regions inside the volume $V_{\rm shock}$ when there are regions with higher than unperturbed pressure $p_1$ and temperature $T_1$, and there are  regions with the lower than unperturbed values, while truly total energy in the volume $V_{\rm shock}$ remains almost the same,  before the shock passage. 
 }
 
 Important comment here is that  
  formula (\ref{initial1})   shows    that there is a finite portion of the volume $V_{\rm shock}$ where temperature is very high 
  $ T\sim 10^8~ {\rm K}$. These regions with high temperature could be 
 the source of    the $10~ {\rm keV}$  x-rays which are normally observed few moments before 
 the flare starts. There is another source of the x-rays in the AQN proposal as mentioned in section \ref{nanoflares}. We elaborate on x ray emission      in next section \ref{observations} devoted to the observational evidences of this proposed mechanism treating AQNs as the triggers of the flares.
 
 Another comment goes as follows. Formulae   (\ref{initial}) and (\ref{initial1}) represent the typical characteristics of the shock waves propagating in the corona.  The nuggets are  not completely annihilated in the corona;  they  continue their journey toward the chromosphere as the typical 
  travelling distance $L$   is sufficiently large according to (\ref{initial}). If these nuggets do not ignite the  large flares they manifest themselves 
 as (observed) ``chromosphere anemone jets"  mentioned at the very end of section \ref{shock}, and to be reviewed  in more details in section \ref{jets}. If a nugget does spark the flare then its  subsequent  evolution can be ignored  because the typical energy associated with fully developed flare is many orders of magnitude larger than a typical 
 initial energy of the AQN.
 
 The second stage of the flare in this framework is represented by  the magnetic reconnection   ignited by the shock wave (characterizing the first stage as described above). We have nothing new to say about this conventional  phase of the evolution. We present the corresponding formula for the total flare's energy for completeness and future estimates, 
 \be
 \label{developed}
 W_{\rm flare}\sim \frac{{\cal{B}}^2}{8\pi}\cdot V_{\rm flare}\sim 0.3\cdot 10^{31} {\rm erg}\left(\frac{{\cal{B}}}{300~ \rm G}\right)^2\cdot\left(\frac{V_{\rm flare}}{10^{13} {\rm km^3}}\right), ~~~ V_{\rm flare}=L_{\perp}^2L_z, 
 \ee
 where $L_z\sim 5000 ~{\rm km}$ is a typical height of the solar corona where the  magnetic field is large, while  $L_{\perp}^2$ is the area in active region (sunspots) which eventually becomes a part of magnetic reconnection producing the large flares. Numerically $L_{\perp}\sim (10^3-10^4)~{\rm km}$ for microflares, and it could be as large as $L_{\perp}\sim 10^5~{\rm km}$ for large flares.  It is assumed that precisely this region of volume $V_{\rm flare}=L_{\perp}^2L_z$ with large average magnetic field $B$ feeds the solar flare as a result of magnetic reconnection. 
 
  It is quite obvious that the energy (\ref{developed}) of a fully developed flare is many orders of magnitude larger  than  the initial energy  related to the shock wave which serves as  a trigger of a large flare. Nevertheless, this  initial  stage in  the flare evolution plays  a key role in future development   of the system  because it provides   a  very strong impulse with $\Delta T/T\gg1$ and $\Delta p^*/p^*\gg1 $ in very small and very localized area for very short period of time (\ref{initial}) in the region where the  magnetic reconnection eventually develops.

  \subsection{\label{scaling}Geometrical  interpretation of the scaling $ {dN} \sim   W^{-\alpha} dW$ }
 In this subsection we would like to interpret the observed scaling (\ref{distribution1}) for the frequency of appearance of {\it large} flares in geometrical terms within  scenario advocated in this work. To avoid confusion with similar scaling  for the {\it nanoflares} one should emphasize from the start of this section that    from the AQN dark matter model perspectives the nanoflares and larger flares
 are considered to be very different types of events. The nanoflares are distributed uniformly through the solar surface and powered   by
 internal energy of the AQNs. In contrast, the large flares are localized in the active regions and powered by   magnetic field as a result of the reconnection. The goal of this subsection is to interpret the scaling (\ref{distribution1}) for large flares, while  the distribution of the nanoflares is governed by completely different physics and it is not the subject of the present studies\footnote{As explained in \cite{Zhitnitsky:2017rop} the corresponding computations from the first principles are hard to carry out as the basic features of the  QCD phase diagram during the formation stage at $\theta\neq 0$ are still unknown. Furthermore,  the time evolution of the AQNs from the formation time until present epoch is also hard to evaluate. A similarity between 
 the observed $\alpha\sim 2$ for large flares and a close  value  for   $\alpha_{\rm nano}\sim 2$ (as some, but not all,  models suggest)  remains a puzzling coincidence at this point as these objects are originated from different physics as emphasized above.}. 
 
 One should emphasize that some discrepancies 
 between different groups (as mentioned at the end of the section  \ref{nanoflares})  on numerical value of the exponent $\alpha_{\rm nano}$ for nanoflare distribution do not affect the present studies  because  our estimates below are based exclusively on the total energy (\ref{total_power}) released by AQNs,  which is consistent with observed EUV luminosity (\ref{estimate}). Therefore, the fundamental relation between nanoflares and flares in the AQN framework is that the nanoflares   play the role of the triggers for large flares as discussed above. Very tiny fraction of these original AQNs  will   ignite  large flares,   while most of the AQNs will heat the corona as reviewed in section \ref{nanoflares}. 
 
 For the  purpose of this work it is more convenient to integrate  formula  (\ref{distribution1}) over energy $dW$ to analyze  the cumulative count
 $ {N} (>W)$ representing total number of flares with energy $W$ and above.
 \be
  \label{distribution2}
   {N} (>W) \simeq   C\left(\frac{W_0}{W}\right) ~~~~~~ {\rm for} ~~~~ \alpha\simeq 2 ~~ {\rm and } ~~W\simeq  (  10^{26}  -  10^{32})~{\rm erg},
  \ee
 where $C$ is the normalization constant. In formula  (\ref{distribution2}) we include energy window for  microflares and flares with their  typical energies. We emphasize that the nanoflares are excluded from this analysis due to the reasons   discussed above. We also assume $\alpha\simeq 2$ as advocated in \cite{Hannah:2007kw}, see also review \cite{Shibata:2016}. The coefficient $C$ is the normalization factor which has dimensionality  ${\rm (year)}^{-1}$  with $W_0\simeq 10^{26} ~{\rm erg}$ as a   part of this normalization's convention. In these notations the  $N(>W)$   represents  the number of flare events per year with energy of order $W$ and above. One can estimate the normalization coefficient $C$  from    known statistics of the flares, see  e.g. Fig.18  in review  \cite{Hannah:2007kw}. In particular,  with our choice 
     $W_0\sim 10^{26} {\rm erg}$ the order of magnitude estimate is  
    $C\sim 10^4 ~{\rm flares/year}$.

 The main  goal of this subsection   is to argue that  the observed $W^{-1}$ scaling in eq.(\ref{distribution2})  can be interpreted in terms of  geometrical parameters of the active regions within our framework. Furthermore, we want to relate the normalization coefficient $C$ with intensity of the observed EUV radiation (\ref{estimate}) which  is determined by the nanoflares (AQN annihilation events). We should emphasize here that we are not computing nor deriving the coefficient $C$ from the first principles in this model. Rather, we want to interpret  the scaling features of eq. (\ref{distribution2}) and the normalization coefficient  $C$  within the AQN framework.

 To accomplish this goal we make  few preliminary comments. 
  A   shock wave (initiated by the AQN) will successfully  ignite  a large flare 
  if it passes through the region where future magnetic reconnection may occur, i.e. 
  the area, where magnetic fluxes have opposite directions and are sufficiently close to each other.   
 From eq. (\ref{eq:SP}) one can infer that a typical distance $l$ between preexisted oppositely directed fluxes   is of order $l\sim L/\sqrt{S}\sim 10~ {\rm m}$ for $S\sim 10^{12}$ and $L\sim 10^4~ {\rm km}$. Our comment here is that the shock front  must pass through  this region $\sim l$  to ignite the large flare. If the shock waves characterized by  parameters (\ref{initial})  do not overlap with this region of pre-existed fluxes   than the corresponding annihilation events will manifest themselves as the  conventional nanoflare type events with sub resolution energies reviewed in Section \ref{nanoflares}. However, if the same AQNs pass through a  small reconnection area $\sim l$ than a large flare may be ignited as will be explained below.

 Our next step is to  estimate 
  the total number of AQNs entering the solar atmosphere per unit time. The corresponding rate can be inferred    from (\ref{total_power}) 
 \be
  \label{nuggets-number}
   N_{\odot \rm (AQN)}\sim \frac{L_{\odot   \rm (AQN)}}{\la {\cal{M}}\ra }\sim  \frac{L_{\odot   \rm (AQN)}}{m_p\la B\ra }\sim 10^{5}\left(\frac{10^{25}}{\la B\ra }\right)\frac{1}{\rm s}\sim 10^{12}\left(\frac{10^{25}}{\la B\ra}\right)\frac{1}{\rm year},  
     \ee  
 which is consistent with the estimates \cite{Benz-2001,Benz-2002} on total number of nanoflares (including the sub resolution events) over the whole Sun, see also related  comments in  \cite{Zhitnitsky:2017rop}.  We should emphasize that we do not specify the mass distribution of the AQNs (or what is the same the nanoflare distribution (\ref{distribution})) in our formula  (\ref{nuggets-number}) by estimating the average number of the annihilation events over entire solar surface averaged over the nuggets mass distribution. This  information is sufficient for the qualitative   estimates    of this work. 

In what follows  we want to understand and interpret  the transition between high rate $N_{\odot \rm (AQN)} \sim 10^{12}$~ {nuggets/year}~representing the total count  (\ref{nuggets-number}) of AQNs entering the Sun and low rate (\ref{distribution2}) for the number of  flares characterized by $C\sim 10^4 ~{\rm flares/year}$  which are ignited by  the same nuggets. We also want to understand/interpret the scaling $W^{-1}$ from (\ref{distribution2}) within AQN scenario. 
  
 The probability that the AQNs fall into the active regions 
 is roughly proportional to the total yearly averaged sunspot area.  The corresponding sunspot area   strongly  depends  on the solar cycle but on average can be estimated as  $\sim 10^4 \mu{\rm Hem}$, see e.g \cite{sunspots}. Therefore, the corresponding suppression factor can be estimated as  $\sim 10^{-2}$. Once again, this should be considered as an order of magnitude estimate   as the fluctuations (during different  years during different   solar   cycles) of the sunspot areas are enormous.  
 
 Another important   suppression factor $\sim (R/L_{\perp})^2$ is 
 related to  the smallness of the  shock wave cone size $R^2$ in comparison with much larger  sunspot area $L_{\perp}^2$
 as explained above.
 It is important to emphasize that the same sunspot area $L_{\perp}^2$ also enters the expression for the volume $V_{\rm flare}=(L_{\perp}^2L_z)$ in eq. (\ref{developed}) which describes the  magnetic  energy    potentially available for   its   transferring  into the flare heating  as a result of magnetic reconnection. 
 We assume in what follows that a typical averaged magnetic field ${\cal{B}}$ over the entire regions  in the active area, the  relevant  height $L_z$ of the solar atmosphere effectively 
 contributing to the total energy of the flare (\ref{developed}) and the shock wave cone size $R^2$  assume (approximately) the same typical values  as these parameters should be treated as the external parameters with respect to  the magnetic reconnection dynamics.
 With the assumption  just formulated we infer that the energy of the flare  scales according to (\ref{developed}) as 
 $W_{\rm flare}\sim L_{\perp}^2$, while   the probability to ignite the magnetic reconnection in area  $L_{\perp}^2$ is proportional to $(R/L_{\perp})^2\sim W^{-1}_{\rm flare}$ as discussed above. Therefore,     the observed relation   (\ref{distribution2}) in our framework has a geometrical interpretation, as  
    the frequency $N(>W_{\rm flare})$ of appearance of a flare with energy of order $W_{\rm flare}$ and above scales as 

\be
\label{scaling1}
N(>W_{\rm flare}) \sim L_{\perp}^{-2}\sim W^{-1}_{\rm flare}. 
\ee
This scaling is consistent with the observed distribution (\ref{distribution2}) and represents a very    important and very generic consequence of the AQN framework.

We are not in position to  compute the normalization constant $C$ from the first principles as we already mentioned, see also comments at the very end of this section. 
  Instead, 
in what follows we would like to interpret a known  normalization factor $C$ in terms of  the AQN proposal: on one side we know the  total number of AQNs entering the solar atmosphere (\ref{nuggets-number}); on the other side we know  the frequency of the flare occurrences (\ref{distribution2}).
We want to relate these two numbers and infer some physical processes which might be responsible for the corresponding suppression factors.

  There are three suppression factors contributing to $C$:
  \be
  \label{C}
  C\equiv C_1C_2C_3 N_{\odot \rm (AQN)} \sim 10^4 ~{\rm flares/year}
  \ee
 The first  two suppression  factors $C_1$ and $C_2$   have been already  mentioned: 
it includes the   smallness of the sunspot areas (suppression $C_1\sim10^{-2}$) and the   smallness of the region which shock wave is capable to sweep being already  inside the active sunspot  area.  The corresponding suppression can be estimated numerically as $C_2\sim R^2/L_{\perp}^2\sim (10^{-4}-10^{-2})$  depending on the specific features of a sunspot region and properties of the nugget,  its trajectory and its relative position with respect to the magnetic configuration (which effectively modify   parameter $R$) in the active spot. In this numerical estimate we use the typical for parameters $R$ magnitude  (\ref{initial}) and  typical   size  $L_{\perp}$  of  the active spots (\ref{developed}).

It is quite obvious that the flare will not start if magnetic configuration in active area is not prepared for the reconnection and  if the   AQN is present in the active spot, but its velocity $v_{\rm AQN}$  is not sufficiently  high such that conditions $M\gg1$ and $M_A>1$ are not satisfied, see footnote \ref{velocity}. 
The corresponding suppression factor  $C_3$   accounts for  this ``preparedness"   of magnetic field configuration
when the reconnection is ready to start if the shock wave sweeps the ``would be" reconnection  region.  
We can  estimate $C_3$ from relation (\ref{C}) by  comparing the observable frequency of appearance of the flares with total number of 
the AQNs entering the solar atmosphere. In particular,  with our normalization convention  one arrives  to the estimate $C_3\simeq (10^{-4}-10^{-2})$.  The corresponding suppression factor can be attributed to  the level of 
 preparedness   of magnetic field configuration for the magnetic reconnection to be successful.  In other words, this suppression factor describes 
 the probability that  
   the magnetic configuration is already built-in and prepared to 
 explode\footnote{In other words, the magnetic fluxes are  oppositely  directed  with a large gradient and properly located  being sufficiently close to each other on a distance of order  $\sim l$.} if the AQN with sufficiently high velocity   is present in the system and  shock wave  passes through this area and triggers the flare.
 
 We conclude this subsection with the following comment. We did not attempt to compute the coefficients $C_i$ from the first principles  
 as such estimates are well beyond the scope of the present work, see also last paragraph of this section and footnotes \ref{MHD} and \ref{MHD_1}
 for comments. Therefore, we cannot provide any uncertainties and error bars in the corresponding estimates. Rather we wanted to understand/interpret 
 the observed difference between large number $N_{\odot \rm (AQN)}$ of nanoflares heating the corona and tiny number of observed large flares
 within our proposal. We formulated the corresponding suppression factors in terms of the dimensionless coefficients $C_i$, which we consider are reasonable. 
 
As we already mentioned,  our original contribution to this field is entirely related to the initial stage of the evolution summarized by parameters   (\ref{initial}),  (\ref{initial1}).  We also argued that the very generic consequence of this proposal  represented by the scaling (\ref{scaling1})  is consistent with the observed rate (\ref{distribution2}). 
Furthermore, this proposal (when the AQNs play the role of the triggers) naturally  resolves  the      problem
of  drastic  separation of  scales when  a flare itself lasts for  about an  hour  while the preparation phase of the magnetic configurations to be reconnected (coded by coefficient $C_3$ discussed above)  could last for  months.  These two  drastically different scales can peacefully coexist in our framework because the presence of a trigger in the system which is not an internal  part of the  magnetic reconnection's  dynamics. 

 We conclude this subsection with the following comment. There are many factors which might be responsible for a slight deviation of the observed scaling with $\alpha\simeq 1.8$ as fitted in ref. \cite{Shibata:2016} and  simple formula (\ref{distribution2}) with $\alpha=2$ which permits   a pure geometrical interpretation as explained above. The corresponding analysis is well beyond the scope of the present work as the main goal here  is to present  a big picture rather than to analyze some minor details  to this proposal.

Explicit numerical simulations supporting the  estimates  presented above in any realistic environment  are  well beyond the scope of the present work due to large number of technical and    conceptual problems which need to be understood\footnote{\label{MHD}In particular,  as we already mention, when the  Mach number $M$
 becomes very large,  the conventional hydro-based computations may not be justified as all relevant parameters (\ref{parameters}) are expressed in terms of the mean free path $l_0$ which obviously inconsistent with conventional treatment of a continuum media when the limit $l_0\rightarrow 0$ is supposed to be smooth. Another    problem with the modelling of this scenario is that the turbulent boundary layer which always accompanies  a  fast moving body should play a key role in the energy exchange of the AQNs with the solar material. This  rate of the energy exchange (which is hard to compute) obviously should play an important role   in any estimates of the dynamics of magnetic reconnection.}.
  Rather, we  present  in next section \ref{observations} some observational evidences supporting  this specific mechanism and 
 entire  framework in general.  
 It is interesting to note  that  a   2d MHD simulations \cite{Tanuma} developed for very different environment with very different purposes  in very different context    nevertheless show that the shock waves indeed may trigger and ignite     sufficiently fast reconnections\footnote{\label{MHD_1}Furthermore, the 2d MHD simulations \cite{Tanuma} show  that  a large number of different phenomena, 
    including SP reconnection \cite{Sweet, Parker1}, Petschek reconnection \cite{Petschek, Kulsrud}, tearing instability, formation of the magnetic islands, and many others, may all take place at different phases in  the evolution of the system, see also reviews \cite{Shibata:2016,Loureiro}.}.

  \section{AQNs as   triggers of solar flares:  proposal   confronting the observations}\label{observations}
  In this section we want to list a number of observations which apparently   show  that the suggested  mechanism 
  when  a large flare is triggered by a small scale jet-like structure due to the shock wave (identified according to  (\ref{identification}) with AQN/nanoflare) is consistent with those observations.
  
  \subsection{Pre-flare x-ray radiation: intensity, timing  and  direction of the flare propagation (from top to bottom)  }
 In this subsection we choose  a specific studies  \cite{x-ray}   of the X-ray analysis  of the X6.9 flare on August 9.2011
in order  to be more specific in our arguments which follow.  We assume that the observed features  are generic properties of   solar flares.  We want to emphasize on three  important elements of the analysis
relevant for our studies  and related to  the initial stage of the evolution,  which is commonly referred  to as a pre-flare phase:

1. It has been observed in \cite{x-ray} that the x-rays in $(0.1-0.8) {\rm nm}$ frequency bands experience very sharp (almost vertical) enhancement with a typical scale of variation measured in seconds. This frequency   band corresponds to $(1.2-10)~ {\rm keV}$ x-rays. 
At the same time,  the $9.4 ~{\rm nm}$ line corresponding to $1~{\rm keV}$ energy shows a less profound, but still noticeable, enhancement  with a typical scale of variation measured in few minutes. Low-energetic $33.5 ~{\rm nm}$ line demonstrates even smaller   variation;

2. The enhancement of the   $(1.2-10)~ {\rm keV}$ band     is enormous and   could be as large as  2-3 orders of magnitude
in comparison with its background values, see Fig.8 in \cite{x-ray};

3.  It has been observed in \cite{x-ray} that ``the pre-flare enhancement propagates from the higher levels of the corona into the lower corona and chromosphere." This claim has been inferred from analysis of intensities of different lines during the pre-flare. 

 In our proposal  all these features are direct consequences of the basic picture and very naturally occur. Indeed,  the high temperature  $T\simeq 10^8$K in the region where the  shock wave propagates is related to  the large Mach number as equation (\ref{initial1}) states. Therefore, it is not a surprise  that the hard x-rays with energy $\sim 10~ {\rm keV}$   can be easily radiated from this region. 
 Furthermore, the very sharp enhancement during very short period of time is also perfectly consistent with our estimates (\ref{initial}) which suggests that precisely this time scale of $(10-10^2)~ {\rm sec} $ determines the typical timing  of the cone produced by  the shock wave.
 \exclude{
 The typical energy associated with this shock wave is estimated in eq. (\ref{initial2}).
 As this energy is released during time $\tau\sim (10-10^2)~ {\rm sec}$ one can infer that on average the corresponding intensity of the radiation  can be estimated as $E_{\rm initial}/\tau\sim (10^{26} -10^{27}){\rm erg/s}$. 
 This energy related to the    x-ray radiation, of course, is many orders of magnitude smaller than typical total energy of a flare (\ref{developed}).
 The corresponding intensity  of the x-ray radiation from the shock wave can be translated in terms of the  flux  measured on the Earth  as follows
 \be
 \label{eq:x-ray}
 \Phi_{\rm shock~ wave }\sim  \left(10^{-3}- 10^{-4}\right) \frac{\rm W}{\rm m^2},
 \ee
 where we assume that the finite portion of the shock wave energy $E_{\rm initial}$ is released as the x-ray radiation. 
   This estimate is perfectly consistent with huge     enhancement of the the $10~ {\rm keV}$ x-rays,  see Fig.8 in \cite{x-ray}.   
 }  
   Finally, the propagation of the flare in the direction from top atmosphere  to its  bottom,  as mentioned in item 3 above, is perfectly consistent with our proposal as the dark matter AQNs which generate the shock waves enter the solar atmosphere from outer space. Therefore, they first enter the higher levels of the corona where they generate the shock wave, before they reach   chromosphere in $\tau\sim (10-10^2)~ {\rm sec}$. 
   
   We should also add that many    related phenomena (such as coronal mass ejection,  proton events, corona darkening  and many others) which often accompany the flares are not very sensitive  to the initial stage of the   flare (which is the topic of the present work) because they are characterized by much larger energy scales. 
  Therefore, the corresponding questions related to a fully developed stage of a flare 
   can  be addressed (and hopefully answered)  within conventional framework 
  such as   MHD  mentioned in footnote \ref{MHD_1} where many  phenomena accompanying the large flares in principle can be studied.

  \subsection{Shapes of the anemone jets}\label{jets} 
  It has been known for quite sometime that the morphology for large scale flares and small scale flares (nanoflares, microflares) are drastically different. To be more specific:  the large scale flares are bubble like or flux rope type, while the small scale flares (nanoflares, microflares) are jets or jet-like,  see e.g. review \cite{Shibata:2016}.
  
   This qualitative difference in the morphology can be easily understood in our framework, where the nanoflares (directly identified  with AQNs according to (\ref{identification})) are always jet like as they inevitably  generate the shock waves as discussed  in Section \ref{flares}. At the same time, if the   AQNs enter  the active regions with large magnetic field, these shock waves  serve as the triggers for the larger flares. In this  case the original morphological shape (jet-like structure) characterized by the  scales (\ref{initial})  is completely washed out by a much larger scale phenomena, the magnetic reconnection, leading to larger flares characterized by drastically different sizes, shapes, and the energy scales (\ref{developed}). 
  
  We want to mention  in this subsection  ref. \cite{Shibata:2007} devoted to analysis  of the anemone jets {\it outside} of sunspots of active regions.
 It has been claimed in   \cite{Shibata:2007} that the observations show the ``ubiquitous presence of chromospheric anemone jets outside of sunspots...".
 The typical characteristics are:
 \be
 \label{eq:jets}
 L\sim (2000-5000)~ {\rm km~ long}, ~~~~~~~ R\sim (150-300)~ {\rm km~ wide}, 
 \ee
 which, in principle,  close to   our expectations  (\ref{initial}) based on the picture where  the ratio $L/R\sim 10$ is large as a result of the development of the shock wave with large Mach number $M\sim 10$ according to (\ref{Mach}). 
  The authors  of  \cite{Shibata:2007}   claim that such jets  represent very generic  events  in {\it quiet} regions which are capable to heat corona and chromosphere.  
  
   However, some precaution should be taken   in any identification 
   of the observed, sufficiently large jets   and tiny jets which must accompany  all the  AQNs. The points is that
   the dominant portion of the nuggets in the allowed window range (\ref{B-range}) are small size AQNs injecting  the energy way  below   the instrumental threshold. Therefore, they could not be directly observed, which makes the identification of AQNs with specific jets   a highly ambiguous procedure. Furthermore, the AQNs could disintegrate  to few pieces when entering the corona, which makes such  identification even more   ambiguous. We do not want to speculate on this topic in the present work as a much better statistics of smaller jets is obviously required for identification of the observed tiny jets.

     \exclude{
  One should comment here that the microflares (in contrast with nanoflares) are   powered by the magnetic field. Nevertheless, they preserve the original    topology of the nanoflares (jet-like) in contrast with large flares  which are bubble like or flux rope like. We think that the basic reason 
  for this difference in topology within AQN scenario is that the magnetic reconnection for the microflares lasts minutes and  follows to the initial trajectory of AQNs. It explains the topological similarity between nanoflares and microflares. At the same time  the  large flares    may last hours, in which case the original information   about initial stage (with 
  jet like structure) of the nanoflares is completely lost  and washed out, and it is entirely determined 
  by the large scale configuration of the magnetic field   as mentioned in footnote \ref{MHD_1}.
  }
  
    \subsection{Other  related phenomena: sunquakes}
    We also want to mention some other observations which might  be also related to the shock waves which are inevitably generated as a result of the dark matter AQNs entering the solar atmosphere from outer space. We want to make few comments 
    on possible relation with sunquakes (acoustic pulses  propagating below the visible surface) which have been observed and analyzed in details, see e.g. recent papers \cite{sunquake,sunquake1}.  However, the nature of the sunquakes remains unknown due to a number of very puzzling features which  can be highlighted   as follows. 
    
    It has been known for quite sometime that the sunquake events  are well correlated with hard x-ray emission (during the impulsive phase) of the flare. The conventional wisdom is  that the flare energy is released in the corona and drives an acoustic disturbance in the solar interior, very close to the  photospheric layer.   For this  mechanism to be operational one should assume that the energy must propagate through nine pressure scale heights. There are many estimates  demonstrating that such energy transfer 
    is almost impossible. At least,   it is very hard to imagine how it could happen   within conventional astrophysical processes \cite{sunquake,sunquake1}. The picture becomes even more puzzling because the  acoustic enhancements occur only in certain locations within the flaring active region. Furthermore,   the  acoustic enhancements   are   not detected for every flare. Authors of ref. \cite{sunquake1} suggested that a single, large temperature increase   might explain some of the observations. However, no any hints (what might be  the cause for such an instantaneous large  temperature increase) were given in \cite{sunquake1}. 
    Authors of ref. \cite{sunquake} suggested that 
``the     energy is transported downwards in a fashion that is somehow {\it invisible} to our observations".

    Our original comment here is that while the energy transport due to the conventional physical processes  is indeed unlikely to occur, 
    the AQNs which play the role of the triggers of the flares as argued in the present work, may easily propagate and penetrate to very deep regions of  the  solar photosphere. After that they can serve as the triggers to ignite the sunquakes. 
     Essentially, we propose that the nuggets   play the same role in photosphere as  they play in upper atmosphere 
     by initiating  the flares as described in this work.     
    
    The main idea behind of this proposal is that a nugget   ignites the flare in the upper atmosphere and continues to move to photosphere without loosing much velocity and momentum as a result of its very small size and very high velocity as reviewed in section \ref{sec:QNDM}. Because the speed of sound at the altitudes  $h\sim ~100 ~{\rm km}$
    could be an order of magnitude smaller than at higher altitudes (\ref{sound}),  the nugget  may ignite   a new  shock wave    in addition to the previous shock wave  in upper corona where it   triggered   the flare as discussed in section \ref{flares}. 
    In other words, estimate  (\ref{delta-pressure}) in photosphere assumes the form
    \be
    \label{Mach1}
    M_{\rm photosphere} &\equiv& \frac{v_{AQN}}{c_s}\simeq 5  \sqrt{ \frac{10^4~ {\rm K}}{T}} \gg 1, ~~~~~~~~ v_{AQN} (h\simeq 100 ~{\rm km})\sim 100~ \frac{\rm km}{\rm s}\\
     \frac{p^*_2}{p^*_1}&\sim &M^2_{\rm photosphere}, ~~~~~ ~~~~~
     \frac{T_2}{T_1}\sim M_{\rm photosphere}^2, \nonumber
    \ee
    such that   shock wave with $M_{\rm photosphere}\gg1$ may trigger the sunquake precisely in the specific location where
   AQN enters a relatively dense region of the photosphere.    This is because the shock wave generated due to the large Mach number  
     may produce a single   highly localized  increase of the  temperature $\Delta T/T\gg 1$ and pressure $\Delta p/p\gg1 $  deep in the atmosphere. 
    
     In this case the observed correlations  between the flares and  the  sunquakes is  a direct  consequence of the fact that both events are ignited by the same quark nugget which moves with velocity $v_{\rm AQN}\sim 100~  {\rm km}/{\rm s}$ when the nuggets reach the photosphere. Furthermore, the observation that  the acoustic enhancements occur only in certain locations within the flaring active region is also easy to understand: it is not the large flare itself which feeds the sunquake. Rather the quake  is localized in the area where the nugget enters the photosphere and ignites a local disturbance leading to the sunquake. The observed larger Doppler velocity shifts in comparison with locations with no sunquake is also 
      easy to interpret and similar to the Doppler shifts discussed for nanoflares as reviewed in Section \ref{nanoflares}
     after eq. (\ref{identification}).

 If we accept this proposal,  one may wonder why    the  acoustic enhancements   are not generated for every flare. The answer  might be related to the fact that in that cases the largest portion of the AQN got annihilated on its way from  corona to photosphere, in which case    a sufficiently   energetic shock wave  could not be formed to play the role of a trigger for the acoustic disturbance.  In other words, only sufficiently large and fast nuggets are capable to reach the photosphere.    
 
 This proposal offers few obvious correlations which, in principle, could be tested with the instruments with sufficiently high resolutions. 
 For example, the x-rays which always accompany the initial stage of the flares should be observed from the same region (with higher altitude) where   the  acoustic enhancements is generated as both effects are originated from the passage of the   same nugget through the atmosphere from corona to photosphere.  Doppler velocity shifts associated with sunquakes must be correlated with similar  Doppler velocity shifts from the same region (with higher altitude) normally associated with nanoflares as reviewed in Section \ref{nanoflares}.   Indeed, both   Doppler velocity shifts  are originated from the passage of the   same nugget through the atmosphere from corona to photosphere. The  Doppler shift at high altitude in the corona signals the initial stage of the flare, while at the photosphere the Doppler velocity shift is associated with the sunquake. 
 
 \section{\label{earth}AQNs impacting  the  Earth   and the Moon}
 
  In this section we want to make few comments on possible tests of the AQN framework for  the nuggets impacting the Earth or the Moon.
  Naively, one could think that the Earth is much simpler system and may  provide much better constraints in comparison with  similar  studies on the Sun.  As we argue below this naive argument,   unfortunately,   is not quite correct.

   The drastic differences  between the two systems (solar atmosphere vs the Earth's atmosphere) within AQN framework have been previously discussed in \cite{Zhitnitsky:2017rop}. 
 From the theoretical viewpoint the solar atmosphere is, in fact,  much simpler system (than the Earth's atmosphere) from AQN perspective.   The basic reason for such simplification is that the solar corona is a highly ionized system consisting  mostly  protons and electrons at high temperature $T\simeq 10^6$ K. It should be contrasted with Earth's atmosphere  where some atoms (mostly heavy elements $N$ and $O$)  are neutral and some are partly ionized. 
 The interaction of these heavy neutral elements  with the AQNs is a highly complicated problem as  the most likely outcome of the collision is the elastic reflection  rather than penetration deep inside the nugget with some probability of partial annihilation processes, which inject the energy and produce the axions.  Furthermore, due to high corona's  temperature, the nuggets become the  electrically charged objects as a result of ionization \cite{Zhitnitsky:2017rop}.  The corresponding enhancement factor in the Sun due to long range Coulomb  forces in highly ionized plasma at temperature $T\simeq 10^6~{\rm K}$ was parametrized 
 in \cite{Raza:2018gpb} by effective size  $R_{\rm eff}\gg R$ to be distinguished from    its physical size $R$. This implies that the effective cross section for protons with AQN in the Sun is approximated as $\sim \pi R_{\rm eff}^2$ while 
 a similar cross section for neutral atoms is much smaller as it can be approximated  as $\sim \pi R^2$.
 Therefore, the annihilation processes are much less efficient in Earth's atmosphere than in the Sun as mentioned above due to the drastic difference between  $R$ and $R_{\rm eff}$ due to the long range Coulomb forces in the ionized hot plasma  \cite{Raza:2018gpb}. 
 
 It has been estimated in \cite{Lawson:2010uz,Lawson:2012vk} that only small portion of the AQN's   mass $\Delta M\simeq 10^{-10} {\rm kg}$ will get annihilated in the Earth's atmosphere. This represents only tiny portion $\sim \Delta B/B \sim 10^{-8}$ of a typical nugget's mass  which can get annihilated in the atmosphere.
On entering the Earth's crust the nugget will continue to deposit energy along 
its path, however this energy is dissipated in the  surrounding rock and is unlikely 
to be directly observable. Generally the nuggets carry sufficient momentum to travel 
directly through the Earth and emerge from the opposite side however a small fraction 
may be captured and deposit all their energy.   In  \cite{Gorham:2012hy} the 
possible contribution of energy deposited by quark nuggets to the Earth's 
thermal budget was estimated and found to be consistent with observations. 
 
 The AQN flux on Earth is estimated as follows
\begin{equation}
\label{eq:flux}
\frac{dN}{dA ~ dt} = n_{\rm AQN}v  \approx 0.3\cdot \left( \frac{10^{25}}{\la B\ra} \right) {\rm km}^{-2} {\rm yr}^{-1}, ~~~~ n_{\rm AQN}\simeq \frac{\rho_{\rm DM}}{m_p \la B\ra}.
\end{equation}
While 
this flux is far below the sensitivity of conventional WIMPs dark 
matter searches it is similar to the flux of cosmic rays 
near the GZK limit.  It has been suggested  that large scale 
cosmic ray detectors 
may be capable of observing  the AQNs  passing through the earth's
atmosphere either through the extensive air shower such an event 
would trigger \cite{Lawson:2010uz} or through the geosynchrotron 
emission generated by the large number of secondary particles
\cite{Lawson:2012vk}, see also \cite{Lawson:2013bya} for review.

It has also been suggested that the \textsc{anita} 
experiment may be sensitive to the radio band 
thermal emission generated by these objects as they pass through the 
antarctic ice \cite{Gorham:2012hy}. These experiments may thus be 
capable of adding direct detection capability to the indirect evidence 
discussed above.  
It has also been estimated in \cite{Abers:2007ji} that, based on Apollo data, 
nuggets of mass from $\sim$ 10 kg to 1 ton (corresponding to 
$B \sim 10^{28\text{-}30}$) must account for less than an order of
magnitude of the local dark matter.  This estimate is perfectly consistent with our window (\ref{B-range}) 
which was based on completely independent analysis.

The observation  of the $E\&M$ radiation  due to the nuggets entering the Earth's atmosphere   requires very large area detectors  
as discussed above.  Furthermore, the  $E\&M$ energy which is dissipated in the  
   mantle or in the Earth's core as a result of the disintegrating of the AQNs in deep underground  will be 
    completely lost for the direct observations. New idea \cite{Fischer:2018niu} which has been advocated recently is that the axions 
   which will be  inevitably produced as a result of the annihilation events in the very deep underground  can be observed. Indeed, 
the corresponding axion flux  can be estimated from (\ref{eq:flux}) as follows \cite{Fischer:2018niu}:
\be
\label{earth-axion}
m_a \Phi ({\rm Earth ~axions})\sim \frac{(2 \Delta B)~{\rm GeV}}{3}\cdot \frac{dN}{dA ~ dt}\sim  10^{16}\cdot \left(\frac{\Delta B}{B}\right)\frac{\rm eV}{\rm cm^2 ~s}, 
\ee
   where  $\Delta B/ B$ is the portion of the   AQNs which  get annihilated in the Earth's core. In formula (\ref{earth-axion}) we assume that   each event of annihilation produces $2~ \rm GeV$ energy deep underground, and  approximately $\sim 1/3$ of this energy is radiated in the form of axions  which will be radiated from  the collapsing axion domain walls surrounding the nuggets.
   We should remind the readers that the axion domain walls    play the key role in the AQN construction    as highlighted in Sect. \ref{sec:QNDM}.
   The energy from the axion domain walls is not available for the observations (similar to the antimatter from antinuggets) unless the AQNs get disintegrated and the energy from axion domain walls get released  in form of  the propagating axions. 
   One should emphasize that  the  axions produced by this mechanism will have the typical velocities $v\sim 0.5 c$ as  computed in \cite{Fischer:2018niu}, in contrast with conventional dark matter axions with typical velocities $v\sim 10^{-3} c$.
   
   Interestingly,   the axion flux  (\ref{earth-axion}) is the same order of magnitude 
   as  the conventional cold dark matter galactic axion contribution. This is because  the parameter 
  ${\Delta B}/{B}\sim 1$ is expected to be order of one because  a finite  portion of the AQNs  will get annihilated in the Earth's core.  However, the 
   wave lengths of  the axions produced due to AQN annihilations, are much shorter  due to their  relativistic velocities  $v\sim 0.5 c$, in contrast with conventional galactic isotropic axions with $v\sim 10^{-3}c$. Therefore, these two distinct contributions can be easily discriminated.  As argued  in  \cite{Fischer:2018niu} 
 the CAST collaboration  which has taken   a significant step forward to become also a haloscope, hopefully, will be capable to discover these axions.

 \section{\label{conclusion} Conclusion and final remarks}

 The main claim of the present work is that the 
 nanoflares conjectured by Parker long ago to resolve the corona heating problem, may also trigger the larger solar flares.
 In our framework the nanoflares are identified with AQNs according to (\ref{identification}). 
  In the previous paper  \cite{Zhitnitsky:2017rop} we argued that the nanoflares may explain the EUV radiation from corona
  as a result of the annihilation processes of the AQNs with the solar material according to  estimate   (\ref{total_power})
  which is perfectly consistent with observations (\ref{estimate}). These order of magnitude estimates have  received a strong numerical support in recent preprint \cite{Raza:2018gpb}. In particular, it has been shown in  \cite{Raza:2018gpb} that the dominant energy injection indeed occurs in vicinity   at altitude around 2000 km, which represents the explanation of the drastic changes in the transition region within AQN framework.  Furthermore, the total luminosity in EUV and soft x-ray bands is computed to be $10^{27} {\rm erg/s}$ in agreement with observations.

  In the present work we argue that exactly the same 
    nuggets entering  the regions with high magnetic field may   ignite  the magnetic reconnections which eventually   lead to much larger flares. This proposal explains   unexpected and bizarre  correlations observed in \cite{Zioutas} between EUV intensity, frequency of flares and positions of the planets as a result of the gravitational lensing of ``invisible" streaming matter towards the Sun.  We identify the  ``invisible" streaming matter from \cite{Zioutas} with AQNs studied in  \cite{Zhitnitsky:2017rop} and in present work. 
    Furthermore, we also argued that the sunquakes may also be triggered by the same AQNs when they reach the photospheric layer.  
  
    Technically, the magnetic reconnection which feeds the flare is ignited due to the shock wave which inevitably develops when the AQN enters the solar atmosphere with sufficiently high velocity,      much greater than  the speed of sound in the corona.
    The corresponding  large value of the Mach number $M=v_{\rm AQN}/c_s\gg 1$ unambiguously implies that the shock waves will be formed and may initiate  the flares due to the  very strong  and very short impulses expressed in terms of  pressure $\Delta p^*/p^*\sim M^2$  and temperature $\Delta T/T\sim M^2$  if occur  in vicinity    of  (would be) magnetic reconnections area in active regions.   
  
  We also argued that this picture is consistent with observations on intensity of the x-ray radiation in pre-flare phase and  studies of the anemone jets. It may   also explain the nature of the   sunquakes   as discussed in section \ref{observations}. It is also consistent with observed scaling for frequency of appearance of a flare with a given energy 
  as argued in section \ref{scaling}.  This proposal (when AQN plays role of a trigger) also resolves a problem of drastic separation of scales  when a flare itself lasts for   an hour while the preparation phase of the magnetic configurations (to be reconnected)  could last for months.

We conclude this work with following proposals   for the futures studies  which may further support (or rule out) this mechanism. \\
  $\bullet$  First of all, a similar EUV and x-ray radiation  correlated with the flares as discussed in the present work (and observed in the Sun, see Fig.\ref{EUV}) must be present 
 in many similar  stars, though the intensity and the spectral properties are highly sensitive to the specific features   of stars and their positions in the galaxy.     The intensity and spectral properties of the radiation must obviously depend on    the  outer dark matter density $\rho_{\rm DM}(r)$ which itself strongly depends on  position of the  star with respect to the galactic center. \\
 $\bullet$ Secondly,     our conjecture that sunquakes are related to the same AQNs which ignite the flares can be tested by
 studies of the cross-correlations similar to analysis  advocated in \cite{Zioutas} (see  Fig. \ref{Earth-Mercury} as a sample) 
 when instead of ``number of flares" one should study the  ``number of sunquakes" along the ``y" axis. \\
  $\bullet$ Thirdly, our arguments  that the large flares are ignited by the   AQNs when they enter the active regions with strong magnetic field can be tested by analyzing  of the cross-correlations similar to studies   advocated in \cite{Zioutas}
  when instead of ``number of flares" one should study the  ``number of sunspots" (or total area of the sunspots) along the ``y" axis
 as a function of the  heliocentric longitude of the planets along ``x" axis. \\
  $\bullet$ Last but not least. The recent proposal  \cite{Fischer:2018niu}  to search for the axions with unusual spectrum with  $v\sim 0.5 c$ 
  might be a smoking gun for this entire framework as it is very hard to imagine how  such  relativistic axions can be   produced by any other mechanism.

\section*{Acknowledgements} 
 I am thankful to Konstantin Zioutas  for explaining  their paper \cite{Zioutas} which  motivated the present studies.   I am also thankful  to Kyle Lawson and   Ludovic Van Waerbeke for discussions and comments.  This research was supported in part by the Natural Sciences and Engineering 
Research Council of Canada.

%\appendix

%\newpage

\end{document}